\documentclass[acmsmall,screen,nonacm]{acmart}
\usepackage[]{collab}
\usepackage{xcolor}
\usepackage{colortbl}
\usepackage{tcolorbox}
\usepackage{mdframed}
\usepackage{listings}
\usepackage{makecell}
\usepackage{multirow}
\usepackage{array}

\usepackage{amsmath,amssymb,amsfonts}
\usepackage{algorithmic}
\usepackage{graphicx}
\usepackage{fbox}
\usepackage{textcomp}
\usepackage{minibox}
\usepackage{balance}
\usepackage{enumitem}
\usepackage{xspace}
\usepackage[T1]{fontenc}
\usepackage{booktabs}
\usepackage{threeparttable}
\usepackage{url}
\usepackage{soul}
\usepackage{ulem}
\usepackage{balance}
\usepackage{subfigure}
\usepackage{diagbox}

\def\BibTeX{{\rm B\kern-.05em{\sc i\kern-.025em b}\kern-.08em
    T\kern-.1667em\lower.7ex\hbox{E}\kern-.125emX}}

\definecolor{ballblue}{rgb}{0.13, 0.67, 0.8}
\definecolor{grey}{rgb}{0.9, 0.9, 0.9}
\definecolor{googlered}{rgb}{0.914, 0.262, 0.207}
\definecolor{dandelion}{rgb}{0.95, 0.65, 0.0}
\definecolor{citecolor}{RGB}{106, 34, 107}

\definecolor{mygreen}{HTML}{AFCFA5}
\newcounter{summary}

\definecolor{mygreen}{HTML}{AFCFA5}
\newcounter{implication}

\definecolor{myyellow}{HTML}{FFF2CC}
\newcounter{finding}

\usepackage[]{collab}
\AtBeginDocument{%
  \providecommand\BibTeX{{%
    Bib\TeX}}}

\setcopyright{none}
\renewcommand\footnotetextcopyrightpermission[1]{}
\usepackage{booktabs}
\usepackage{tabularx}
\usepackage{pifont}
\newcommand\nm{SWEOps-Bench\xspace}

\newcommand{\eg}{{\em e.g.},\xspace}
\newcommand\num{55\xspace}

\begin{document}


\title{SWEOps-Bench: Evaluating Agents for End-to-End Resolution of Software-Induced Cloud Incidents}

\title{SWEOps-Bench: Evaluating Agents for End-to-End Resolution of Software-Induced Cloud Incidents}

\author{Danwen Chen}
\affiliation{%
  \institution{Sun Yat-sen University}
  \city{Guangzhou}
  \country{China}
}
\email{chendw9@mail2.sysu.edu.cn}

\author{Guangba Yu}
\affiliation{%
  \institution{The Chinese University of Hong Kong}
  \city{Hong Kong}
  \country{China}
}
\email{guangbayu@cuhk.edu.hk}

\author{Pengfei Chen}
\affiliation{%
  \institution{Sun Yat-sen University}
  \city{Guangzhou}
  \country{China}
}
\email{chenpf7@mail.sysu.edu.cn}

\renewcommand{\shortauthors}{Chen et al.}



\begin{abstract}
As autonomous agents advance, the traditional boundary between software development and cloud operations gradually blurs, increasing the need for integrated incident detection, localization, and remediation.
Existing benchmarks evaluate isolated diagnostic queries or repository-level program repair, failing to test continuous incident resolution across operational clusters and source repositories.
We introduce \nm, an interactive benchmark comprising \num reproducible incident tasks derived from confirmed historical defects across 16 Kubernetes operator projects.
Each task preserves the original software repository and initializes in a healthy cluster before a controlled workload triggers the historical fault.
Agents investigate runtime evidence and submit source or configuration repairs through a terminal interface, while the harness builds and deploys the changes within a shared time budget.

Evaluation covers stage-wise accuracy, verified recovery, penalized lifecycle latency, and token consumption. Across six agent–model configurations, faulty-state detection ranges from 72.7\% to 90.9\%, whereas overall recovery ranges from only 14.5\% to 52.7\%, exposing a substantial gap between recognizing incidents and resolving them end to end. Trajectory analysis further shows that reported diagnosis and final recovery only partially coincide: some tasks achieve verified recovery despite an incorrect reported RCA, while runtime symptom clearance can still fail fix-specific functional verification. \nm establishes a rigorous foundation for evaluating autonomous agents in integrated development and operational environments.

\end{abstract}



\maketitle

\section{Introduction}
\label{sec:introduction}

Modern cloud systems underpin critical digital services, and incidents can disrupt service availability and impose substantial operational costs~\cite{incident2020,Incident2022SoCC,alertguardian2025ASE}.
When incidents originate from software defects, resolving them requires both operational investigation and changes to the underlying implementation.
LLM-based agents that inspect running systems and modify source code create an opportunity to integrate development and operations within a continuous incident-resolution workflow.
Kubernetes operators provide a representative setting for this integration, where software defects directly impair managed cloud workloads~\cite{Acto2023SOSP,OAT2026NSDI}.
This resolution lifecycle couples runtime anomaly detection, root cause analysis (RCA), and remediation.
A central question is whether autonomous agents can resolve software-induced cloud incidents end to end within practical time.

\begin{figure*}[t]
    \centering
    \includegraphics[width=\textwidth]{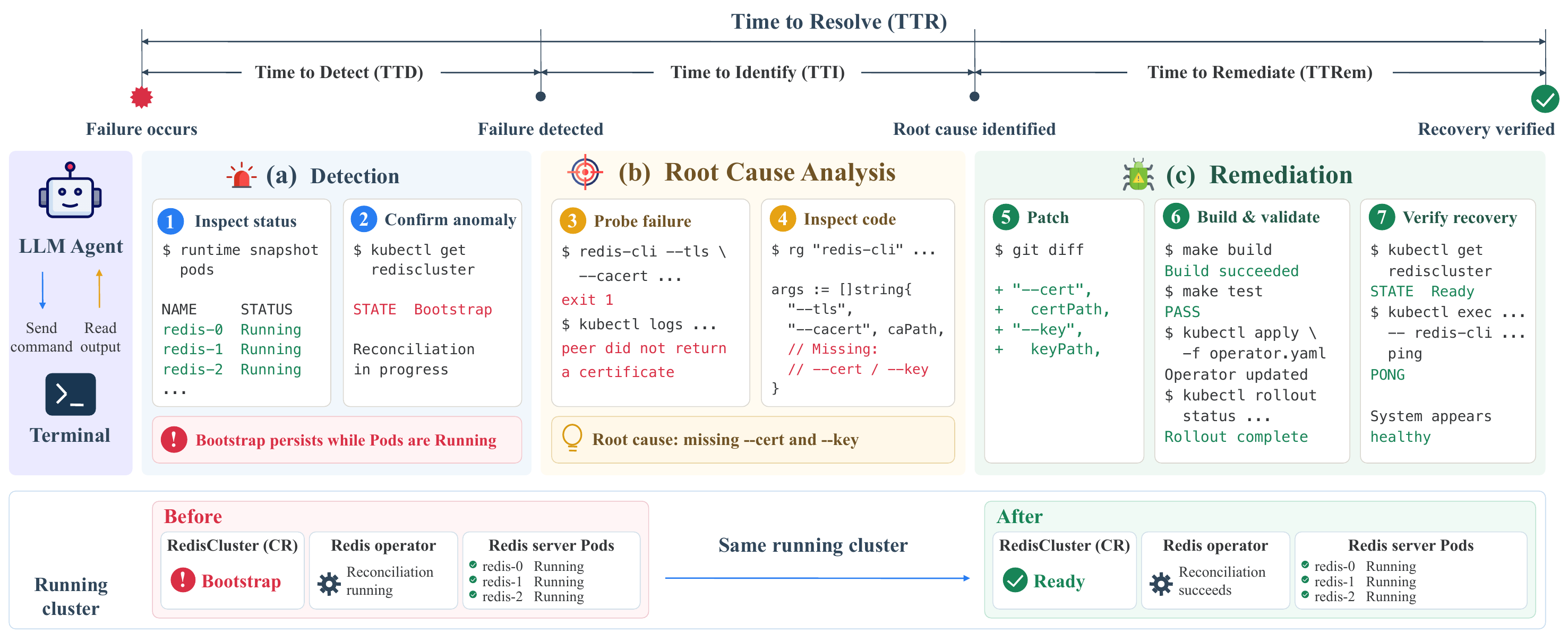}
    \vspace{-0.2in}
\caption{An example incident resolution lifecycle, spanning Time to Detect (TTD), Time to Identify (TTI), and Time to Remediate (TTRem) toward the overall Time to Resolve (TTR).}
    \label{fig:lifecycle}
      \vspace{-0.1in}
\end{figure*}

To illustrate this integrated resolution process, Figure~\ref{fig:lifecycle} depicts a Redis operator incident caused by missing TLS client certificate arguments~\cite{redisOperator1864}.
When Redis requires client certificates, the operator repeatedly fails cluster health checks, preventing the \texttt{RedisCluster} resource from reaching the \texttt{Ready} state despite healthy server Pods.
During detection, the agent inspects runtime cluster observations to identify this operational stall.
During root-cause analysis, it combines operator logs with client probes to trace the TLS authentication failure to missing \texttt{-{}-cert} and \texttt{-{}-key} arguments in \texttt{redis-cli} invocations.
During remediation, the agent modifies the controller implementation, while the harness builds and deploys the repair and returns runtime feedback. Recovery is then verified through fault clearance and functional checks.
This case underscores that resolving realistic cloud incidents requires traversing the complete lifecycle rather than inspecting code or telemetry in isolation.

Existing benchmarks evaluate isolated segments of the incident resolution lifecycle, as summarized in Table~\ref{tab:benchmark_comparison}.
Cloud operations benchmarks evaluate anomaly detection and operational mitigation, predominantly relying on synthetic fault injection in microservices~\cite{chen2025aiopslab,ITBench2025ICML,OpenRCA2025ICLR,RCA2025Eval,wang2026cloudopsbench}.
However, they do not require agents to repair historical software defects in their original repositories or validate deployed recovery.
Conversely, software engineering benchmarks evaluate program repair from static issue descriptions against offline test suites~\cite{Defects4J,SWEBench2024ICLR,yang2025swebench}.
These protocols omit runtime incident discovery in live clusters and operational recovery verification.
Consequently, existing benchmarks leave open whether autonomous agents can navigate the complete incident resolution lifecycle to achieve verified recovery.

Addressing these limitations requires an evaluation setting grounded in confirmed historical defects and live operational execution.
Reproducing incidents within their original repositories preserves authentic failure modes, triggering conditions, and native build workflows.
In real-world cloud environments, resolution efficiency is paramount: delayed anomaly detection and failed repair attempts directly prolong service disruptions~\cite{li2022incident,Incident2022SoCC}.
Evaluation must therefore treat time not merely as an operational timeout, but as an intrinsic performance objective across the lifecycle~\cite{ITILIncidentManagement,LogicMonitorMTTR}.
This requires assessing verified recovery alongside the operational latency (\eg Time to Identify (TTI) and Time to Resolve (TTR)) and resource expenditures necessary to achieve it.

To realize these requirements, we introduce \nm, an interactive benchmark for evaluating agents on end-to-end resolution of software-induced cloud incidents.
\nm contains \num incident tasks reproduced from confirmed historical defects in 16 open-source Kubernetes operator projects managing databases, stream processors, and infrastructure services.
Each task initializes in a healthy cluster environment to evaluate false-positive resistance before an automated trigger induces the target incident.
Agents interact with the live cluster and source repository through a bash terminal, while the benchmark harness builds and deploys candidate repairs to return runtime execution feedback.
Across this lifecycle, \nm evaluates both the effectiveness and efficiency of autonomous agents in achieving verified recovery.

We evaluate terminal-based agent harnesses, including Codex and Claude Code, across four recent proprietary and open-weight models spanning GPT-5.6 Luna, Claude-Sonnet-5, DeepSeek-V4.1-Flash, and GLM-5.3-Flash.
Our evaluation characterizes incident resolution effectiveness alongside lifecycle latency and token consumption, revealing three central empirical findings.
First, end-to-end resolution remains substantially harder than incident recognition: faulty-state detection ranges from 72.7\% to 90.9\%, while overall recovery ranges from 14.5\% to 52.7\% across the evaluated configurations. Second, relative advantages shift across lifecycle stages and agent–model configurations, and substantially greater token consumption does not consistently translate into higher recovery. Third, diagnosis and remediation are only partially aligned: for the two representative configurations examined at the trajectory level, 14.5\%--21.8\% of tasks achieve verified recovery despite an incorrect reported joint RCA, while runtime symptom clearance can still violate fix-specific functional requirements.

Our principal contributions are summarized as follows:
\begin{itemize}[leftmargin=*]
\item \textbf{Realistic Benchmark.}
We construct \nm with \num incident tasks derived from confirmed historical defects in 16 open-source projects.
Each task preserves the original repository context and reproduces the failure in an isolated cluster environment.
\item \textbf{Evaluation Framework.}
We formulate end-to-end incident resolution as an interactive task spanning runtime detection, source code-level diagnosis, and deployed source repair.
The framework combines curated ground truth with executable recovery checks to evaluate resolution effectiveness and efficiency.
\item \textbf{Empirical study.}
We evaluate frontier terminal-based agent harnesses and models on \nm, characterizing resolution capabilities, stage bottlenecks, and resource expenditures.
\end{itemize}

\section{Background and Motivation}
\label{sec:background}

\subsection{Cloud Application Management}
\label{sec:background_management}

Kubernetes operators automate application management by combining custom resources with dedicated software controllers~\cite{Acto2023SOSP,OAT2026NSDI}.
Operators encode domain-specific operational expertise into software and interact with the platform API server to manage application workloads.
These controllers execute control loops governed by the state-reconciliation principle: they continuously observe the actual state of managed resources, compare it with the desired state declared in the custom specification, and issue operations to drive the system toward agreement~\cite{Sieve2022OSDI}.

The reliability of management software is critical because defects in operator logic affect the stability of the managed applications (\eg by generating invalid resource specifications or preventing managed applications from reaching their desired states)~\cite{Acto2023SOSP,OAT2026NSDI}.
For example, a defect in a Redis operator causes the generated Pod specification to include an init container with a volume mount that references an undefined volume when persistence is disabled~\cite{redisOperator1540}.
The Kubernetes API server rejects the resulting Pod specification, preventing the cluster from completing initialization.
The failure manifests in runtime resources, whereas the underlying defect resides in the operator implementation.
Resolving such incidents therefore requires connecting runtime symptoms to the responsible controller source code, repairing the defect, and verifying the recovery.

\subsection{Cloud Incident Resolution Lifecycle}
\label{sec:background_lifecycle}

In production environments, resolving software-induced incidents is not an isolated, one-off action, but an operational lifecycle governed by dependability and Site Reliability Engineering (SRE) principles~\cite{Avizienis2004Taxonomy,ITILIncidentManagement,AtlassianIncidentResponse}.
Prior empirical studies show that cloud incident management spans continuous phases of anomaly detection, diagnostic investigation, and mitigation~\cite{li2022incident,Incident2022SoCC}.
In practice, these activities form an interdependent feedback loop: diagnostic hypotheses guide candidate repairs, while runtime execution feedback from deployed changes informs further investigation and code revision.
Because prolonged disruptions severely degrade service availability, industrial practice treats resolution latency as a primary operational metric alongside functional restoration~\cite{LogicMonitorMTTR}.

As illustrated in Figure~\ref{fig:lifecycle}, \nm models this resolution lifecycle across three foundational stages:
\begin{itemize}[leftmargin=*]
\item \textbf{Detection.}
Detection concerns recognizing abnormal system behavior from runtime observations.
In cloud-native environments, resource states, platform events, and controller logs provide complementary evidence, as individual resource status does not necessarily establish overall application health.

\item \textbf{Root Cause Analysis.}
RCA traces an observed runtime failure to its underlying cause.
For software-induced incidents, source-level localization identifies the files and functions, or configuration files, responsible for the failure.

\item \textbf{Remediation.}
Remediation addresses the underlying defect to restore required system functionality.
For implementation bugs, this involves modifying and redeploying the affected software, followed by executable verification confirming both failure elimination and preservation of existing capabilities.
\end{itemize}

Together, these stages capture the progression from incident onset to verified recovery.
We characterize lifecycle efficiency across this progression using Time to Detect (TTD), Time to Identify (TTI), Time to Remediate (TTRem), and overall Time to Resolve (TTR), whose formal boundaries are defined in Section~\ref{sec:evaluation-methodology}.

\subsection{Prior Benchmarks and Motivation}
\label{sec:prior_benchmarks_motivation}

\begin{table*}[t]
\centering
\caption{Comparison of \nm with related benchmarks.}
\vspace{-0.1in}
\label{tab:benchmark_comparison}
\footnotesize
\setlength{\tabcolsep}{2.2pt}
\renewcommand{\arraystretch}{1.18}

\providecommand{\cmark}{\textcolor{green!60!black}{\ding{51}}}
\providecommand{\xmark}{\textcolor{red!70!black}{\ding{55}}}

\begin{tabularx}{\textwidth}{@{}
    l
    >{\hsize=1.16\hsize\centering\arraybackslash}X
    >{\hsize=0.84\hsize\centering\arraybackslash}X
    c c c c c
@{}}
\toprule
\shortstack{\textbf{Benchmark}}
& \shortstack{\textbf{Target}\\\textbf{Systems}}
& \shortstack{\textbf{Task}\\\textbf{Construction}}
& \shortstack{\textbf{Detect}\\\textbf{Eval}}
& \shortstack{\textbf{RCA}\\\textbf{Eval}}
& \shortstack{\textbf{Source}\\\textbf{Repair}}
& \shortstack{\textbf{Runtime}\\\textbf{Recovery}}
& \shortstack{\textbf{Full}\\\textbf{Lifecycle}} \\
\midrule
OpsEval~\cite{OpsEval2025FSE}
& Curated scenarios & Case questions & \xmark & \cmark* & \xmark & \xmark & \xmark \\

AIOpsLab~\cite{chen2025aiopslab}
& Demo microservices & Fault injection & \cmark & \cmark & \xmark & \cmark & \xmark \\

OpenRCA~\cite{OpenRCA2025ICLR}
& Enterprise telemetry & Failure records & \xmark & \cmark & \xmark & \xmark & \xmark \\

ITBench (SRE)~\cite{ITBench2025ICML}
& Demo microservices & Fault injection & \xmark & \cmark & \xmark & \cmark & \xmark \\

SREGym~\cite{SREGym}
& Demo microservices & Fault injection & \xmark & \cmark & \xmark & \cmark & \xmark \\

Cloud-OpsBench~\cite{wang2026cloudopsbench}
& Demo microservices & Fault injection & \xmark & \cmark & \xmark & \xmark & \xmark \\

RCAEval~\cite{RCA2025Eval}
& Demo microservices & Fault injection & \xmark & \cmark & \xmark & \xmark & \xmark \\

SWE-bench~\cite{SWEBench2024ICLR}
& Open-source repos & Curated issues & \xmark & \xmark & \cmark & \xmark & \xmark \\

\midrule
\textbf{\nm}
& \textbf{Open-source operators} & \textbf{Historical defects} & \cmark & \cmark & \cmark & \cmark & \cmark \\
\bottomrule
\end{tabularx}
\vspace{1mm}
\raggedright
\scriptsize{\textit{Notes:} 
*OpsEval assesses diagnostic reasoning through curated Q\&A. 

\textbf{Full Lifecycle} connects detection, diagnosis, source repair, and recovery verification within one evaluation protocol, measuring elapsed time from incident onset to the recovery declaration for subsequently verified tasks.
}
\vspace{-0.2in}
\end{table*}

As autonomous agents increasingly bridge system operations and code modification, incident handling shifts from fragmented tasks toward unified workflows.
However, as summarized in Table~\ref{tab:benchmark_comparison}, existing benchmarks evaluate isolated segments of this lifecycle under disparate formulations.
Cloud operations benchmarks assess diagnostic reasoning and operational mitigation, typically relying on synthetic fault injection in microservices~\cite{chen2025aiopslab,ITBench2025ICML,wang2026cloudopsbench,SREGym} or offline telemetry snapshots~\cite{OpenRCA2025ICLR,nezha2023fse,RCA2025Eval,OpsEval2025FSE}.
However, they do not require agents to resolve defects in the original controller codebase or validate recovery after redeployment.
Conversely, software engineering benchmarks evaluate repository-level program repair from static issue descriptions~\cite{SWEBench2024ICLR,yang2025swebench}, omitting runtime incident discovery and deployed operational validation.

To evaluate autonomous agents across the complete incident resolution lifecycle, the research community requires an interactive benchmark grounded in confirmed historical defects reproduced within their original repositories.
Reproducing authentic defects captures complex reconciliation failures while providing developer patches and regression suites as objective ground truth.
These considerations motivate four essential design requirements for lifecycle evaluation:
\begin{itemize}[leftmargin=*]
\item \textbf{R1: Runtime incident discovery.}
Evaluation requires agents to detect anomalies directly from live cluster observations without pre-packaged issue descriptions.
Agents independently discriminate healthy baseline states from emerging failures before initiating diagnostic investigation.

\item \textbf{R2: Deployable source repair.}
Resolving software-induced incidents requires correcting the defective implementation rather than applying temporary operational workarounds.
Candidate repairs should integrate into the native build workflows and deploy to the live cluster, enabling feedback-driven iterative revision.

\item \textbf{R3: Verified functional recovery.}
Symptom disappearance alone does not establish resolution, as superficial patches or pod restarts can mask failures while introducing regressions.
Verification combines case-specific fault oracles with end-to-end management checks to confirm that the defect is eliminated and required functionality is preserved.

\item \textbf{R4: End-to-End lifecycle evaluation.}
Rather than evaluating isolated sub-tasks, evaluation spans detection, diagnosis, source repair, and recovery verification under standardized operational budgets. This unified protocol exposes how agents backtrack from deployment failures and the overall latency and token costs incurred.
\end{itemize}

\begin{figure*}[t]
    \centering
    \includegraphics[width=\textwidth]{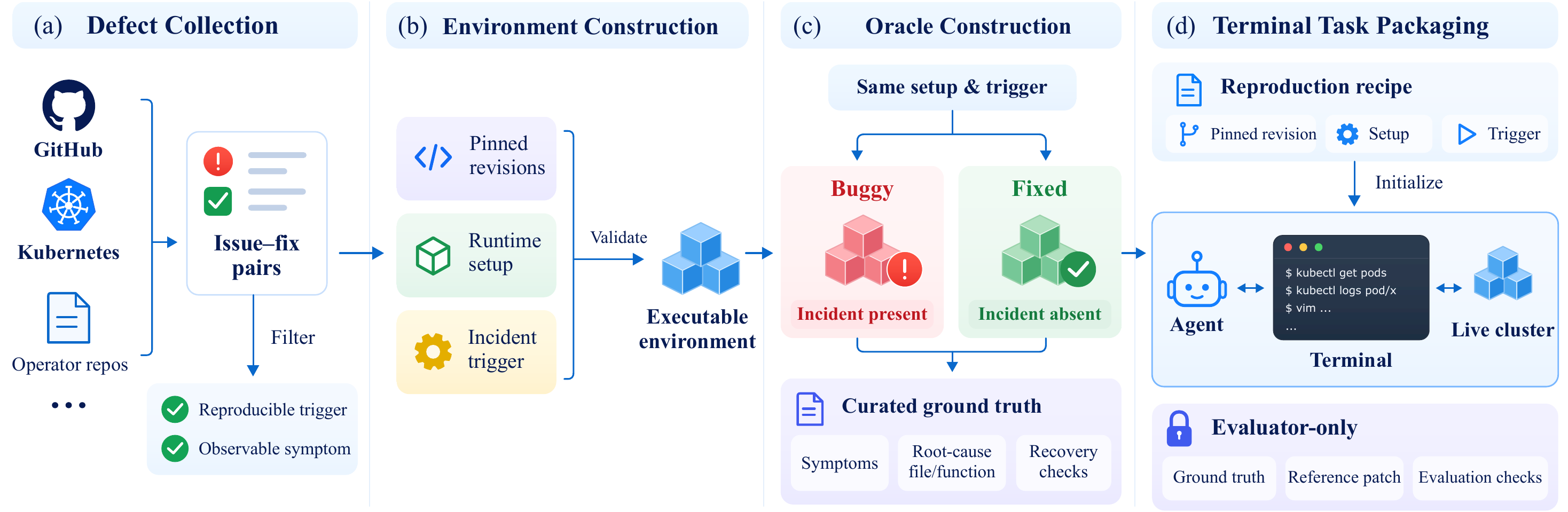}
    \vspace{-0.2in}
    \caption{Overview of the \nm task construction pipeline.}
    \vspace{-0.2in}
    \label{fig:overview}
\end{figure*}
\section{SWEOps-Bench Construction}

To realize the design requirements established in Section~\ref{sec:prior_benchmarks_motivation}, we present \nm, an interactive benchmark for end-to-end autonomous incident resolution.
Figure~\ref{fig:overview} summarizes its construction pipeline, from historical defect collection and environment reconstruction to oracle construction and terminal-task packaging.
Sections~\ref{sec:defect-collection}--\ref{sec:evaluation-methodology} detail the dataset construction, evaluation oracles, task execution protocol, and metrics for resolution effectiveness, lifecycle efficiency, and token consumption.

\subsection{Task Formulation}
\label{sec:task-formulation}


\nm formulates incident resolution as an interactive task over a live Kubernetes environment and its management software repository. The agent receives runtime observations and terminal access, and produces anomaly decisions, a root-cause report, source or configuration changes, and a recovery declaration.

\begin{table*}[t]
\centering
\caption{Upstream operator projects and incident distributions in \nm.}
\vspace{-0.1in}
\label{tab:operator_projects}
\footnotesize
\setlength{\tabcolsep}{2pt}
\renewcommand{\arraystretch}{1.18}

\begin{tabularx}{\textwidth}{@{}
    l
    >{\raggedright\arraybackslash}X
    >{\centering\arraybackslash}p{0.95cm}
    >{\centering\arraybackslash}p{0.85cm}
    >{\centering\arraybackslash}p{0.75cm}
    !{\hspace{0.4em}\vrule width 0.6pt\hspace{0.4em}}
    l
    >{\raggedright\arraybackslash}X
    >{\centering\arraybackslash}p{0.95cm}
    >{\centering\arraybackslash}p{0.85cm}
    >{\centering\arraybackslash}p{0.75cm}
@{}}
\toprule
\makecell[l]{\textbf{Operator}} &
\makecell[c]{\textbf{Managed}\\\textbf{System}} &
\makecell[c]{\textbf{Stars}} &
\makecell[c]{\textbf{KLOC}} &
\makecell[c]{\textbf{Cases}} &
\makecell[l]{\textbf{Operator}} &
\makecell[c]{\textbf{Managed}\\\textbf{System}} &
\makecell[c]{\textbf{Stars}} &
\makecell[c]{\textbf{KLOC}} &
\makecell[c]{\textbf{Cases}} \\
\midrule
\href{https://github.com/OT-CONTAINER-KIT/redis-operator}{RedisOp} & Redis & 1,464 & 10.1 & 9 &
\href{https://github.com/pravega/zookeeper-operator}{ZKOp} & ZooKeeper & 383 & 2.5 & 3 \\

\href{https://github.com/VictoriaMetrics/operator}{VMOp} & VictoriaMetrics & 590 & 50.6 & 9 &
\href{https://github.com/grafana/grafana-operator}{GrafanaOp} & Grafana & 1,389 & 11.5 & 2 \\

\href{https://github.com/mongodb/mongodb-kubernetes-operator}{MongoOp} & MongoDB & 1,357 & 10.1 & 5 &
\href{https://github.com/kubeflow/spark-operator}{SparkOp} & Apache Spark & 3,154 & 11.4 & 2 \\

\href{https://github.com/percona/percona-server-mongodb-operator}{PCN/MongoOp} & MongoDB & 457 & 27.9 & 5 &
\href{https://github.com/open-telemetry/opentelemetry-operator}{OTelOp} & OpenTelemetry & 1,762 & 28.4 & 2 \\

\href{https://github.com/cloudnative-pg/cloudnative-pg}{CNPG} & PostgreSQL & 9,374 & 62.4 & 4 &
\href{https://github.com/rook/rook}{Rook} & Ceph & 13,670 & 61.1 & 2 \\

\href{https://github.com/k8ssandra/cass-operator}{CassOp} & Cassandra & 228 & 13.1 & 3 &
\href{https://github.com/kube-logging/logging-operator}{LoggingOp} & Fluent Bit & 1,715 & 19.6 & 1 \\

\href{https://github.com/knative/operator}{KnativeOp} & Knative & 225 & 6.0 & 3 &
\href{https://github.com/prometheus-operator/prometheus-operator}{PromOp} & Prometheus & 9,983 & 33.7 & 1 \\

\href{https://github.com/percona/percona-xtradb-cluster-operator}{PCN/XtraDBOp} & Percona XtraDB & 625 & 25.8 & 3 &
\href{https://github.com/rabbitmq/cluster-operator}{RabbitMQOp} & RabbitMQ & 1,117 & 5.8 & 1 \\
\midrule
\multicolumn{10}{@{}l@{}}{\textbf{Total: 16 operator repositories \quad \quad \num\ historical incident tasks}} \\
\bottomrule
\end{tabularx}

\vspace{1.5mm}
\raggedright
\scriptsize
\textit{Notes:} Operator names provide clickable hyperlinks directly to their upstream GitHub repositories. Stars and KLOC were measured on September 29, 2026.
\vspace{-0.1in}
\end{table*}

\subsection{Dataset Construction}
\label{sec:dataset-construction}
\nm contains \num incident tasks constructed from confirmed historical defects in real-world Kubernetes management software.
Below, we detail our candidate defect collection, two-phase screening methodology, and execution environment reconstruction.

\subsubsection{Defect Collection and Selection}
\label{sec:defect-collection}
Following controller reliability studies~\cite{Acto2023SOSP,OAT2026NSDI}, we select operator projects based on project maturity, source code availability, and application diversity.
As summarized in Table~\ref{tab:operator_projects}, the benchmark encompasses 16 representative operator projects across diverse functional domains, including distributed databases, stream processing, messaging, observability, and serverless. All repositories  are officially developed and maintained by recognized open-source communities (\eg CNCF) or reputable commercial software vendors (\eg MongoDB).

We collect candidate defects from prior controller reliability studies and GitHub issue histories.
The former include defects documented by Sieve~\cite{Sieve2022OSDI}, Acto~\cite{Acto2023SOSP}, and Oat~\cite{OAT2026NSDI}.
To reflect modern cloud-native operational challenges and reduce the potential pre-training exposure window, the second source mines contemporary bug fixes merged between January and August 2026 across 16 repositories, inspecting maintainer-confirmed pull requests with bug-related tags on the main development branch.

We screen candidates using three exclusion criteria:
(E1) \emph{Unverified fix}: no verifiable merged upstream fix is available, or the referenced fix was reverted;
(E2) \emph{Out-of-scope defect}: the change is confined to documentation, comments, third-party deployment tooling, or the managed application rather than the operator implementation or operator-owned configuration;
(E3) \emph{Duplication or reconstruction constraints}: the report duplicates a defect already represented by the same upstream fix, or reproduction requires unavailable proprietary services or infrastructure.
Candidates that pass this screening proceed to environment reconstruction and execution-based validation.

\subsubsection{Environment Reconstruction and Validation}
\label{sec:environment-reconstruction}
For each candidate defect passing initial screening, we reconstruct an isolated runtime environment using the original management software (Figure~\ref{fig:overview}(b)).
We pin the affected source revision, Kubernetes release, container images, dependencies, and deployment manifests used to reproduce the target failure.
The environment must reach a healthy initial state in which the operator and managed resources satisfy the required health checks.
A case-specific triggering procedure then exposes the existing defect without modifying the target implementation.

We validate each reconstructed case through paired executions of the buggy and upstream-fixed revisions under the same environment and triggering procedure.
This execution-based screening excludes candidates that meet any of the following criteria:
(E4) \emph{Bootstrapping failure}: the cluster, operator, or managed resources cannot reach the required healthy initial state;
(E5) \emph{Unobservable failure}: the defect has no verifiable runtime manifestation in Kubernetes resources, management operations, or managed workloads;
(E6) \emph{Unreproducible environment or trigger}: required dependencies cannot be obtained, or the triggering procedure cannot reliably reproduce the target failure.
A task is retained only if the buggy revision reproduces the target incident and the upstream-fixed revision eliminates the failure while passing the required functional and regression checks.
This two-phase screening process yields \num verified incident tasks: 28 from prior controller reliability studies and 27 from the additional 2026 collection.

\begin{figure*}[t]
    \centering
    \includegraphics[width=\textwidth]{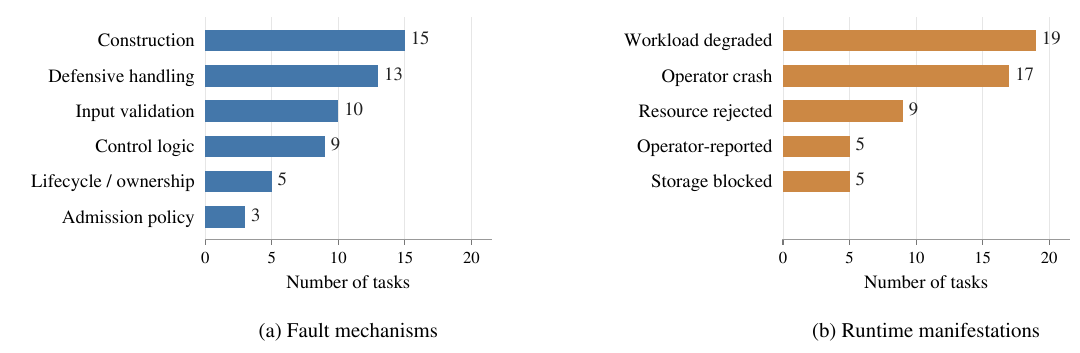}
    \vspace{-0.3in}
    \caption{Distribution of \num incident tasks by (a) primary fault mechanism and (b) dominant runtime manifestation. Bar-end labels indicate task counts.}
    \vspace{-0.2in}
    \label{fig:composition}
\end{figure*}

Figure~\ref{fig:composition} summarizes the distribution of the final \num incident tasks.
Across defect mechanisms (Figure~\ref{fig:composition}(a)), the tasks are grouped into six categories: Construction (15 cases), Defensive Handling (13), Input Validation (10), Control Logic (9), Lifecycle / Ownership (5), and Admission Policy (3).
Across runtime manifestations (Figure~\ref{fig:composition}(b)), the tasks are grouped into five categories: Workload degraded (19 cases), Operator crash (17), Resource rejected (9), Operator-reported (5), and Storage blocked (5).

\subsection{Evaluation Oracle Construction}
\label{sec:ground-truth-and-oracles}

To evaluate autonomous agents across the complete lifecycle, \nm constructs dedicated evaluation oracles for each operational stage (Figure~\ref{fig:overview}(c)).

For runtime detection, we construct an executable fault oracle implemented as a case-specific probe derived from the differential behavior between buggy and fixed revisions.
Depending on the failure mechanism, the probe checks Kubernetes resource conditions, controller event logs, or workload service endpoints for evidence of the target incident.
Observations are labeled as normal when the system satisfies the required health checks and as abnormal when the fault oracle detects the target failure.
The earliest confirmed manifestation establishes the reference timestamp for measuring Time to Detect (TTD).

Reference root-cause annotations are derived from maintainer-confirmed issue reports and merged upstream fixes.
We inspect the affected implementation and reproduced runtime behavior to verify the defect mechanism and identify the responsible source or configuration files and, where applicable, functions.
Auxiliary changes in the fixing commit are not automatically treated as root-cause locations.

For remediation, each task incorporates executable checks for build and deployment success, target fault elimination, and required functional behavior.
The fault oracle determines whether the target failure remains observable in the running cluster.
To verify genuine recovery, fix-specific functional tests exercise the affected capability under triggering conditions, while regression suites check selected previously working behavior.
These test oracles are constructed by adapting upstream PR regression tests and project integration suites, confirmed by differential execution across the buggy and fixed revisions.
Crucially, these checks evaluate the runtime behavior of the repaired implementation rather than textual similarity to the upstream developer patch, accommodating alternative valid repairs while ensuring symptom elimination does not come at the expense of required functionality.

\subsection{Task Packaging and Execution}
\label{sec:task-packaging}

Each incident task is packaged into an execution bundle comprising the environment specification, pinned operator repository, automated triggering workload, and hidden evaluation oracles (Figure~\ref{fig:overview}(d)).
To restrict access to external task information, outbound internet access is disabled within the agent's tool execution environment, restricting network connectivity strictly to the local Kubernetes cluster.
The original issue discussions, reference fixes, and verification scripts remain under benchmark control and are withheld from the agent throughout execution.

The benchmark harness equips the agent with an interactive bash terminal providing direct command-line access to the live cluster via \texttt{kubectl} and Unix utilities, alongside read-write access to the local operator repository.
Agents use terminal commands to inspect system behavior, search and modify source files.
When a candidate repair is submitted, the harness compiles the source code, rebuilds the controller image, deploys the updated controller into the cluster, and returns execution feedback, such as compiler errors or rollout failures.
This feedback loop enables iterative revision, with file modifications, runtime logs, and cluster state changes carrying forward continuously within the interactive session.

Each evaluation session proceeds through a standardized execution protocol.
The harness first validates cluster readiness and presents the agent with an initial observation from a healthy baseline state, serving as a negative control to evaluate false-positive detection decisions.
The healthy-state phase operates under a separate wall-clock budget of $B = 60$ minutes.
The harness then executes the case-specific triggering workload, and a benchmark-controlled fault probe establishes when the target incident becomes observable.
Once the incident manifests, the fault-state phase starts with a fresh conversation context and operates under an independent wall-clock budget of $B = 60$ minutes covering all subsequent investigation, diagnosis, code modification, and deployment attempts.
The fault-state phase ends when the agent explicitly declares recovery in the terminal or when the operational budget expires.
The harness then applies the hidden repair and recovery checks to the submitted changes and resulting system state.

\subsection{Evaluation Methodology and Metrics}
\label{sec:evaluation-methodology}

\nm evaluates autonomous incident resolution across stage-wise decision quality, verified remediation, lifecycle latency, and resource costs.
Evaluation builds directly upon the ground-truth annotations and executable oracles established in Section~\ref{sec:ground-truth-and-oracles}. Agent implementations may internally invoke specialized subagents for detection, RCA, and remediation; such delegation is treated as an internal orchestration mechanism and does not alter the evaluation protocol or metrics.

To evaluate anomaly detection, we assess whether agents reliably discriminate healthy operational states from active incidents.
\emph{Healthy} measures correct recognition during the initial negative-control baseline, and \emph{Faulty} measures correct anomaly identification once the target incident manifests.
We report \emph{Paired} detection, which strictly requires both decisions to be correct within the same task execution.
This paired metric explicitly penalizes trivial heuristics that indiscriminately report perpetual normality or continuous anomalies.

We evaluate RCA at the \emph{file} and \emph{function} levels and additionally report \emph{joint accuracy}, which requires both locations to be correct.
For configuration-induced incidents where function-level ground truth does not apply, joint correctness evaluates whether the causal configuration files are accurately identified.
Detection and diagnosis outcomes are reported independently from final system recovery, enabling empirical characterization of instances where agents achieve verified recovery despite incomplete or incorrect diagnostic reports.

To dissect remediation, Table~\ref{tab:remediation_outcomes} reports pass rates for seven outcomes: code compilation (\emph{Build}), container rollout (\emph{Deploy}), a non-empty, parseable patch modifying a ground-truth root-cause file (\emph{Patch Check}), fault clearance confirmed by the fault probe (\emph{Fault Cleared}), regression testing (\emph{Regression Tests}),  verification of affected capabilities under triggering conditions (\emph{Fix-Specific Tests}), and \emph{Agent-Declared Recovery}, which indicates whether the agent explicitly declares recovery within the execution budget. \emph{Patch Check} is auxiliary and not a recovery criterion, since a valid repair may modify alternative implementation locations. Pass rates are computed over applicable tasks, counting unattempted or failed checks as unsuccessful. Overall \emph{Recovery} requires an explicit recovery declaration within the execution budget and success on all applicable criteria except \emph{Patch Check}.

To evaluate lifecycle efficiency, we measure Time to Detect (\emph{TTD}), Time to Identify (\emph{TTI}), Time to Remediate (\emph{TTRem}), and overall Time to Resolve (\emph{TTR}) in minutes.
Reporting latency only for successful executions can obscure differences in completion rates and in the sets of tasks solved.
For task $i$, detection duration $T_{i,\mathrm{D}}$ is measured from incident manifestation to the first correct anomaly report.
Identification duration $T_{i,\mathrm{I}}$ is measured from the first correct anomaly report to the first correct root-cause report.
Remediation duration $T_{i,\mathrm{Rem}}$ is measured from the first correct root-cause report to the recovery declaration, provided that recovery is subsequently verified. If no correct root-cause report precedes verified recovery, we instead use the most recent RCA report before the recovery declaration as the starting point.
A stage duration is defined only when both boundary milestones occur in the specified order.
These durations measure intervals between reporting milestones rather than time spent exclusively on each activity.
Overall resolution time is measured from incident manifestation to the recovery declaration for tasks that subsequently pass recovery verification.
It remains defined even when an intermediate reporting milestone is missing.

Inspired by penalized runtime evaluation~\cite{PAR}, we retain observed durations for achieved
milestones and assign predetermined penalties otherwise.
An unachieved detection, identification, or remediation milestone receives a penalty of
$B$, $3B$, or $6B$ minutes, respectively.
Let $\widetilde{T}_{i,\mathrm{D}}$, $\widetilde{T}_{i,\mathrm{I}}$, and
$\widetilde{T}_{i,\mathrm{Rem}}$ denote these penalized stage latencies.

For the overall resolution, we retain the observed end-to-end TTR, measured to the agent's recovery declaration, when task $i$ subsequently passes recovery verification. Otherwise, we aggregate the penalized stage latencies:
\begin{equation}
\widetilde{\mathrm{TTR}}_i =
\begin{cases}
T_{i,\mathrm{TTR}}, & \text{if verified recovery is achieved},\\
\widetilde{T}_{i,\mathrm{D}} +
\widetilde{T}_{i,\mathrm{I}} +
\widetilde{T}_{i,\mathrm{Rem}}, & \text{otherwise}.
\end{cases}
\end{equation}

Thus, a successfully recovered task retains its observed wall-clock resolution time even if an
intermediate reporting milestone is missing, whereas unresolved tasks are penalized according to
the lifecycle stages they fail to complete.
A task achieving none of the milestones receives the maximum combined penalty of
$10B$ minutes.
These penalized quantities are evaluation metrics expressed in time units and do not extend the session budget.
We report mean penalized TTD, TTI, TTRem, and TTR across all \num benchmark tasks.

Finally, we measure operational overhead by reporting mean input and output token counts per trajectory separately for healthy-state and fault-state execution.
Input counts include cached tokens, and we additionally report the cached-input proportion (\emph{Cache (\%)}).

\section{Empirical Evaluation}
\label{sec:empirical-evaluation}

We evaluate LLM-based agents on \nm through four research questions on lifecycle effectiveness, efficiency, robustness, and trajectory behavior:

\textbf{RQ1 (Lifecycle Effectiveness):} How effectively do agents perform at individual lifecycle stages and achieve verified incident resolution?

\textbf{RQ2 (Lifecycle Efficiency):} What time and token costs do agents incur across the incident resolution lifecycle?

\textbf{RQ3 (Robustness):}
How robust is agent resolution effectiveness across incident characteristics?

\textbf{RQ4 (Trajectory Analysis):} What trajectory behaviors shape incident resolution outcomes?

\begin{table*}[t]
    \centering
    \caption{Overall performance of evaluated agent--model
    configurations on SWEOps-Bench.}
    \label{tab:overall_performance}
    \vspace{-0.1in}
    \small
    \setlength{\tabcolsep}{3.5pt}
    \renewcommand{\arraystretch}{1.1}

    \begin{tabular}{llccccccc}
        \toprule
        \multirow{2}{*}{Agent}
        & \multirow{2}{*}{Model}
        & \multicolumn{3}{c}{Detection (\%)}
        & \multicolumn{3}{c}{RCA (\%)}
        & \multirow{2}{*}{REC (\%)} \\

        \cmidrule(lr){3-5}
        \cmidrule(lr){6-8}

        & 
        & Healthy 
        & Faulty
        & Paired
        & File
        & Func. 
        & Joint
        & \\

        \midrule

        \multirow{3}{*}{Codex}
        & GLM-5.3-Flash
        & 72.7 & 74.5 & 60.0 & 54.5 & 43.6 & 43.6 & 34.5 \\

        & DeepSeek-V4.1-Flash
        & 43.6 & 85.5 & 38.2 & 63.6 & \textbf{50.9} & \textbf{50.9} & 45.5 \\

        & GPT-5.6 Luna
        & 85.5 & 76.4 & 72.7 & 50.9 & 36.4 & 36.4 & 30.9 \\

        \midrule

         \multirow{3}{*}{Claude Code}
        & GLM-5.3-Flash
        & 67.3 & 72.7 & 45.5 & 25.5 & 5.5 & 5.5 & 14.5 \\

        & DeepSeek-V4.1-Flash
        & 83.6 & 87.3 & 72.7 & \textbf{65.5} &  30.9 & 30.9 & 47.3 \\

        & Claude-Sonnet-5
        & \textbf{94.5} & \textbf{90.9} & \textbf{85.5} & \textbf{65.5} & 38.2 & 38.2 & \textbf{52.7} \\

        \bottomrule
    \end{tabular}
    \vspace{-0.2in}
\end{table*}

\subsection{Experimental Setup}
\label{sec:experimental-setup}

All evaluations follow the protocol and metrics defined in Sections~\ref{sec:task-packaging}--\ref{sec:evaluation-methodology}.

\subsubsection{Agents and Models}
We evaluate two terminal-based agent harnesses: Codex and Claude Code.
We evaluate four models spanning proprietary and open-weight models available during the evaluation period.
Specifically, our proprietary suite comprises OpenAI GPT-5.6 Luna (\texttt{gpt-5.6-luna}, 2026-07-09) and Anthropic Claude-Sonnet-5 (\texttt{claude-sonnet-5}, 2026-06-30).
Our open-weight suite comprises DeepSeek-V4.1-Flash (2026-09-10) and Zhipu GLM-5.3-Flash (\texttt{glm-5.3-flash}, 2026-08-26).
All evaluations directly query the official first-party API endpoints provided by the respective model vendors between August and September 2026, avoiding third-party proxies.
We employ the official default generation parameters and reasoning-effort configurations across all models without bespoke hyperparameter tuning.

\subsubsection{Execution Infrastructure}
\label{sec:harbor}
We use the Harbor benchmarking framework~\cite{Harbor} to orchestrate and manage containerized task environments across \nm.
For each incident task, Harbor provisions an isolated Kubernetes cluster along with the corresponding management operator and runtime dependencies.
Each cluster operates with pinned Kubernetes releases matching the historical defect requirements, drawn from four versions: v1.21.14+k3s1, v1.23.17+k3s1, v1.28.10+k3s1, and v1.31.5+k3s1. All task environments run on a single virtual machine provisioned through
OpenStack Nova with KVM virtualization, configured with 16 vCPUs, 64 GiB of
memory, and Ubuntu 24.04.4 LTS (x86\_64). Each task environment is capped at
2 vCPUs and 7 GiB of memory, with the memory limit increased to 10--12 GiB for
four memory-intensive cases. To limit resource contention, we run at most four
task environments concurrently.

We conduct one independent run per task and agent--model configuration.
Harbor captures full interaction trajectories, shell command logs, runtime resource snapshots, and token consumption metrics for post-execution analysis.

\subsection{RQ1: Lifecycle Effectiveness}
\label{sec:rq1}

\begin{table*}[t]
    \centering
    \caption{Detailed remediation outcomes of evaluated
    agent--model configurations. All results are reported as percentages. Component rates use applicable-task denominators.}
    \label{tab:remediation_outcomes}
    \vspace{-0.1in}
    \small
    \setlength{\tabcolsep}{3.5pt}
    \renewcommand{\arraystretch}{1.1}

    \begin{tabular}{llccccccc}
        \toprule

        Agent
        & Model
        & Build
        & Deploy
        & \shortstack{Patch\\Check}
        & \shortstack{Fault\\Cleared}
        & \shortstack{Regression\\Tests}
        & \shortstack{Fix-Specific\\Tests}
        & \shortstack{Agent-Declared\\Recovery} \\

        \midrule

         \multirow{3}{*}{Codex}
        & GLM-5.3-Flash
        & 100 & 63.6 & 56.4 & 50.9 & 100 & 29.4 & 47.3 \\

        & DeepSeek-V4.1-Flash
        & 100 & \textbf{89.1} & 72.7 & 63.6 & 100 & 47.1 & 60.0 \\

        & GPT-5.6 Luna
        & 100 & 76.4 & 58.2 & 54.5 & 100 & 35.3  & 47.3 \\

        \midrule

        \multirow{3}{*}{Claude Code}
        & GLM-5.3-Flash
        & 100 & 29.1 & 27.3 & 25.5 & 97.9 & 35.3 & 20.0 \\

        & DeepSeek-V4.1-Flash
        & 100 & 87.3 & \textbf{80.0} & \textbf{72.7} & 97.9 & 41.2 & 60.0 \\

        & Claude-Sonnet-5
        & 100 & 81.8 & 72.7 & 61.8 & 97.9 & \textbf{58.8} & 72.7 \\

        \bottomrule
    \end{tabular}
    \vspace{-0.1in}
\end{table*}

Table~\ref{tab:overall_performance} reports effectiveness across detection, RCA, and overall recovery (REC). Agents generally recognize active incidents more reliably than they discriminate between healthy and faulty states. Faulty-state detection ranges from 72.7\% to 90.9\%, while paired detection ranges more widely from 38.2\% to 85.5\%. The largest gap appears for Codex + DeepSeek-V4.1-Flash, which achieves 85.5\% faulty-state detection but only 38.2\% paired accuracy. By contrast, Claude Code + Claude-Sonnet-5 achieves the highest paired detection rate of 85.5\%. These results demonstrate the importance of evaluating healthy and faulty states jointly: high anomaly recognition alone does not imply reliable incident detection.

Performance becomes more differentiated from detection to source-level diagnosis and recovery. File-level RCA ranges from 25.5\% to 65.5\%, while function-level and joint RCA range from 5.5\% to 50.9\%. All configurations show lower function-level than file-level accuracy, indicating that identifying the relevant file is more reliable than pinpointing the responsible implementation. Relative model performance also changes across lifecycle stages. Under Codex, GPT-5.6 Luna achieves the highest paired detection, whereas DeepSeek-V4.1-Flash achieves the highest joint RCA and recovery. Under Claude Code, Claude-Sonnet-5 leads paired detection, joint RCA, and recovery. Thus, stronger detection or RCA does not consistently translate into stronger end-to-end resolution.

The harness comparison shows a similarly non-uniform pattern for the two models evaluated with both agents. With GLM-5.3-Flash, Codex performs better in paired detection, joint RCA, and recovery. With DeepSeek-V4.1-Flash, Claude Code achieves higher paired detection and recovery, while Codex achieves higher joint RCA. These contrasting results indicate that neither harness provides a uniform advantage across models and lifecycle stages. Agent effectiveness should therefore be interpreted at the agent--model configuration level rather than attributed to the model or harness in isolation.

Table~\ref{tab:remediation_outcomes} provides a fine-grained view of the remediation process, from build and deployment to runtime fault clearance, functional verification, and agent-declared recovery. Build succeeds on 100\% of applicable tasks, and broad regression-test pass rates remain between 97.9\% and 100\%, while deployment and fix-specific verification are substantially less reliable. Runtime symptom clearance does not establish semantic repair correctness: a modification can remove the immediate failure while changing how the operator handles supported requests or invalid inputs. Fix-specific verification examines these management requirements beyond the disappearance of the triggering symptom.

\begin{figure*}[t]
    \centering
    \vspace{-0.1in}\includegraphics[width=\textwidth]{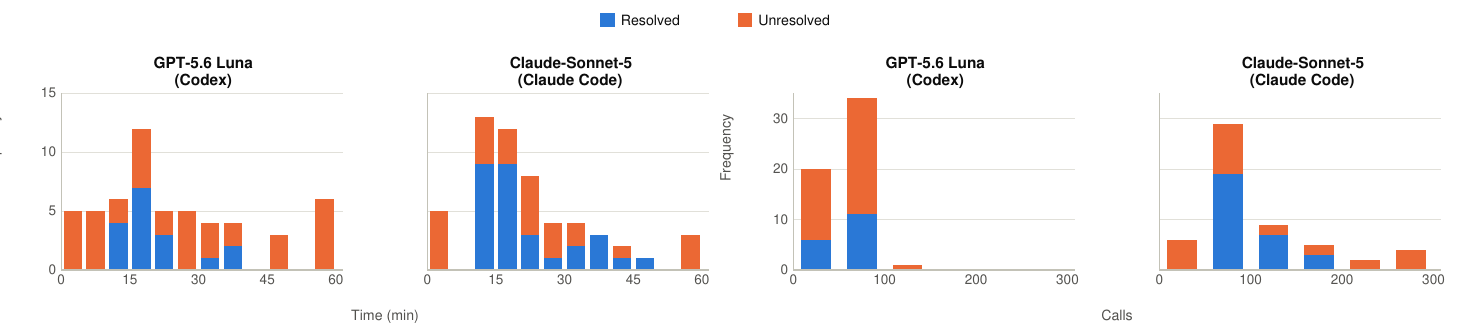}
    \caption{Distributions of agent execution time and number of model calls.
}
    \vspace{-0.1in}\label{fig:execution_distribution}
\end{figure*}

\begin{table*}[t]
    \centering
    \caption{Penalized lifecycle latencies of evaluated agent--model configurations (min),
    averaged over all 55 tasks.}
    \label{tab:lifecycle-latency}
    \small
    \setlength{\tabcolsep}{6pt}
    \renewcommand{\arraystretch}{1.1}

    \begin{tabular}{llrrrr}
        \toprule
        Agent & Model
        & $\widetilde{\mathrm{TTD}}$
        & $\widetilde{\mathrm{TTI}}$
        & $\widetilde{\mathrm{TTRem}}$
        & $\widetilde{\mathrm{TTR}}$ \\
        \midrule

        \multirow[c]{3}{*}{Codex}
        & GLM-5.3-Flash
        & 21.2 & 103.5 & 239.6 & 351.7 \\

        & DeepSeek-V4.1-Flash
        & 17.4 & 96.2 & 208.0 & 285.7 \\

        & GPT-5.6 Luna
         &19.7 & 117.4 & 252.3 & 361.9  \\

        \midrule

        \multirow[c]{3}{*}{Claude Code}
        & GLM-5.3-Flash
        & 23.1 & 171.1 & 309.4 & 477.4 \\

        & DeepSeek-V4.1-Flash
         &13.3 & 129.1 & 194.3 & 269.8  \\

        & Claude-Sonnet-5
        & 10.9 & 116.1 & 177.3 & 263.7 \\

        \bottomrule
    \end{tabular}
\end{table*}

\subsection{RQ2: Lifecycle Efficiency}
\label{sec:rq2}


Table~\ref{tab:lifecycle-latency} reports penalized lifecycle latencies. Penalized latencies increase substantially toward later stages, with TTD consistently lower than TTI and TTRem, indicating that lifecycle cost is concentrated after incident recognition.

Penalized TTR broadly tracks recovery performance, although configurations with comparable TTR can still exhibit different recovery rates. For example, Claude Code + Claude-Sonnet-5 and Claude Code + DeepSeek-V4.1-Flash achieve penalized TTRs of 263.7 and 269.8 minutes, with recovery rates of 52.7\% and 47.3\%, respectively. Figure~\ref{fig:execution_distribution} further shows that unresolved trajectories can persist despite extended execution and many model calls. Thus, penalized TTR is informative but not a one-to-one proxy for recovery, while greater execution effort does not uniformly translate into successful recovery.


Fault-state token consumption differs substantially across configurations (Table~\ref{tab:token-consumption}). For the two models evaluated with both harnesses, Claude Code uses 4.1--4.5$\times$ more input tokens and 6.6--9.6$\times$ more output tokens than Codex, yet recovery decreases for GLM-5.3-Flash (34.5\% to 14.5\%) and increases only slightly for DeepSeek-V4.1-Flash (45.5\% to 47.3\%). Thus, greater token use does not by itself imply greater end-to-end effectiveness.


\subsection{RQ3: Robustness across Incident Characteristics}
\label{sec:rq3}

\begin{figure*}[t]
    \centering
    \includegraphics[width=\textwidth]{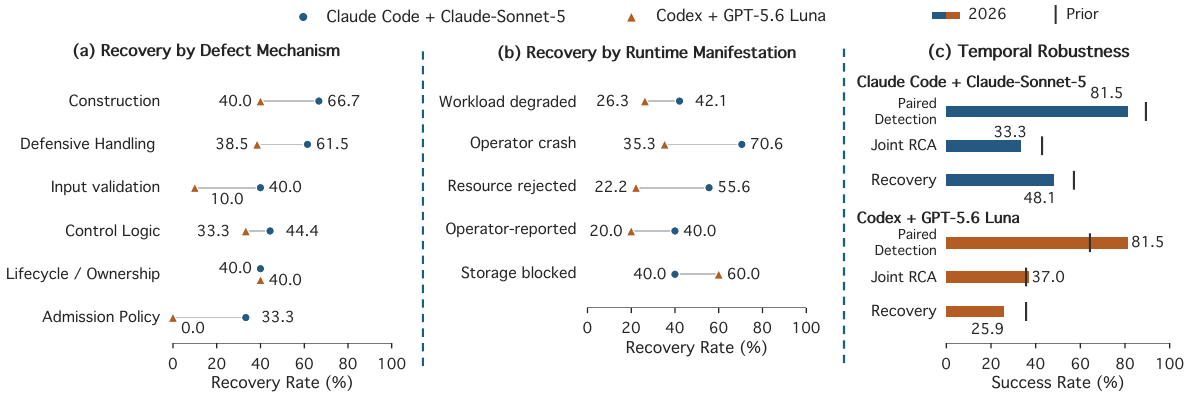}
    \caption{Agent effectiveness across incident characteristics: (a) recovery by defect mechanism, (b) recovery by runtime manifestation, and (c) lifecycle-stage performance by issue recency. 
}
    \label{fig:incident characteristics}
    \vspace{-0.1in}
\end{figure*}

For focused analysis, we select one vendor-native proprietary configuration for each agent harness: Codex + GPT-5.6 Luna and Claude Code + Claude-Sonnet-5. Figure~\ref{fig:incident characteristics} examines whether agent effectiveness is robust across variations in defect mechanism, runtime manifestation, and issue recency.

Recovery varies substantially across defect mechanisms and runtime manifestations (Figure~5(a--b)).
Input Validation constitutes a shared weakness, with recovery rates of 40.0\% for Claude Code + Claude-Sonnet-5 and 10.0\% for Codex + GPT-5.6 Luna.
These tasks require agents to distinguish supported from invalid inputs.
Normalization or fallback handling can resolve valid cases but may also suppress validation errors for inputs that should instead be rejected.
For example, one agent normalized an invalid version value and incorrectly treated the resulting error clearance as success.

Beyond this common weakness, the configurations exhibit different patterns across categories.
Claude Code + Claude-Sonnet-5 achieves higher recovery in Defensive Handling (61.5\% vs. 38.5\%) and Operator crash (70.6\% vs. 35.3\%), whereas Codex + GPT-5.6 Luna achieves higher recovery in Storage blocked (60.0\% vs. 40.0\%).
The two configurations tie on Lifecycle / Ownership (40.0\%).
These differences show that aggregate recovery rates mask substantial variation across incident characteristics and agent configurations.
\begin{table*}[t]
    \centering
    \caption{Token consumption of evaluated agent--model configurations,
    separated by healthy-state and fault-triggered execution.
    Values are means per trajectory. Cached tokens are included in input tokens.}
    \label{tab:token-consumption}
    \vspace{-0.1in}
    \small
    \setlength{\tabcolsep}{4pt}
    \renewcommand{\arraystretch}{1.1}

    \begin{tabular}{llrrrrrr}
        \toprule
        & &
        \multicolumn{3}{c}{Healthy State} &
        \multicolumn{3}{c}{Fault State} \\
        \cmidrule(lr){3-5}
        \cmidrule(lr){6-8}

        Agent
        & Model
        & Input
        & Output
        & Cache (\%)
        & Input
        & Output
        & Cache (\%) \\
        \midrule

        \multirow[c]{3}{*}{Codex}
        & GLM-5.3-Flash
        & 292.8K &  5.1K & 82.9
        & 910.4K & 15.2K &  89.2 \\

        & DeepSeek-V4.1-Flash
        & 3127.5K &24.0K  & 98.5
        &  1902.2K & 21.5K  & 97.8 \\

        & GPT-5.6 Luna
        & 654.8K    & 7.4K    & 86.8
        & 1845.7K  & 19.8K   & 88.8 \\

        \midrule

        \multirow[c]{3}{*}{Claude Code}
        & GLM-5.3-Flash
        & 853.5K & 23.9K & 83.4
        & 3703.3K & 100.8K & 79.8 \\
        
        & DeepSeek-V4.1-Flash
        & 2536.8K & 66.3K &  95.6
        & 8500.4K & 206.8K & 96.1 \\

        & Claude-Sonnet-5
        & 669.4K & 7.5K & 78.7
        & 4620.6K & 44.1K & 75.9\\

        \bottomrule
    \end{tabular}
    \vspace{-0.1in}
\end{table*}

We further examine performance by issue recency because older incidents have had more opportunity to appear in publicly available data. Figure~\ref{fig:incident characteristics}(c) compares the 28 incidents from prior studies with the 27 collected in 2026. Claude Code + Claude-Sonnet-5 performs better on prior incidents across all three metrics, whereas Codex + GPT-5.6 Luna performs better on 2026 incidents for paired detection and joint RCA but worse for recovery. The largest recency gap is 17.2 percentage points in paired detection for Codex + GPT-5.6 Luna, while Claude Code + Claude-Sonnet-5 shows smaller differences across the three metrics. Overall, the results show no uniform degradation on newly collected incidents.



\subsection{RQ4: Trajectory Behaviors and Resolution Outcomes}
\label{sec:rq4}
\begin{figure*}[t]
    \centering
    \vspace{-0.2in}\includegraphics[width=\textwidth]{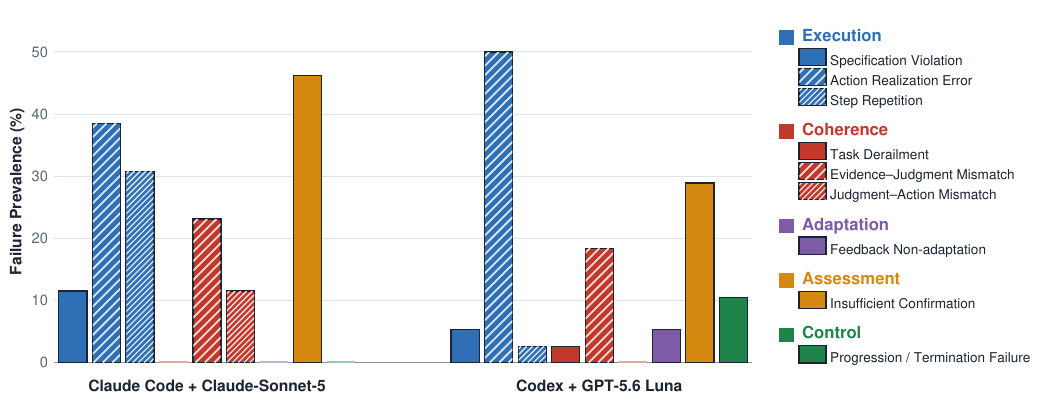}
    \caption{Prevalence of behavioral failure modes in failed trajectories across the two agent configurations.
}
    \label{fig:failure_modes}
    \vspace{-0.1in}
\end{figure*}

\begin{figure*}[t]
    \centering
    \includegraphics[width=\textwidth]{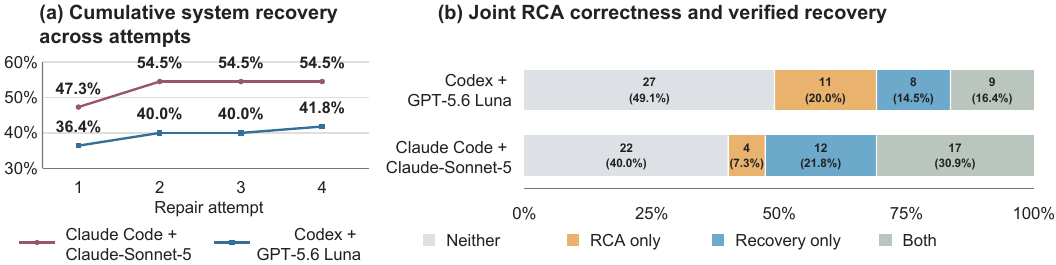}
    \caption{(a) Cumulative system-side recovery across repair attempts; (b) relationship between joint RCA correctness and Overall Recovery.
}
    \label{fig:rca_recovery}
    \vspace{-0.1in}
\end{figure*}
To understand how trajectory-level behaviors shape incident resolution outcomes, we further analyze these two representative configurations.  We first manually inspect failed trajectories using a taxonomy adapted from the trajectory analysis of Terminal-Bench~\cite{terminalbench}. Two annotators calibrated and refined the category definitions on a subset of trajectories and then annotated the failed trajectories, resolving disagreements through discussion. The taxonomy covers execution, coherence, adaptation, assessment, and control. Annotations are multi-label, so a trajectory may exhibit multiple failure modes. Figure~\ref{fig:failure_modes} compares the prevalence of these modes across the two configurations. We then examine when verified recovery occurs across repair attempts and how reported RCA correctness corresponds to final verified recovery, before using representative trajectories to illustrate how these patterns arise during execution.

As shown in Figure~\ref{fig:failure_modes}, failures in both configurations are concentrated in Action Realization Error (ARE) and Insufficient Confirmation (IC). For Claude Code + Claude-Sonnet-5, IC occurs in 46.2\% of failed trajectories and ARE in 38.5\%; for Codex + GPT-5.6 Luna, the corresponding rates are 28.9\% and 50.0\%. Claude Code additionally exhibits substantial Step Repetition (30.8\%), whereas Codex shows more Evidence--Judgment Mismatch (EJM, 18.4\%). Other modes are less frequent. Judgment--Action Mismatch (JAM) occurs only in Claude Code (11.5\%), while Task Derailment, Feedback Non-adaptation, and Progression / Termination Failure are absent for Claude Code + Claude-Sonnet-5 and remain at or below 10.5\% for Codex + GPT-5.6 Luna.

Action Realization Error often reflects incomplete implementation of a plausible intervention, such as modifying only some locations implicated in a defect. Insufficient Confirmation and Evidence--Judgment Mismatch instead arise when agents observe residual anomalies but dismiss them on insufficient evidence. The key issue is therefore not whether an anomaly is observed, but whether its exclusion from the resolution target is justified.

\begin{figure*}[t]
    \centering
    \includegraphics[width=\textwidth]{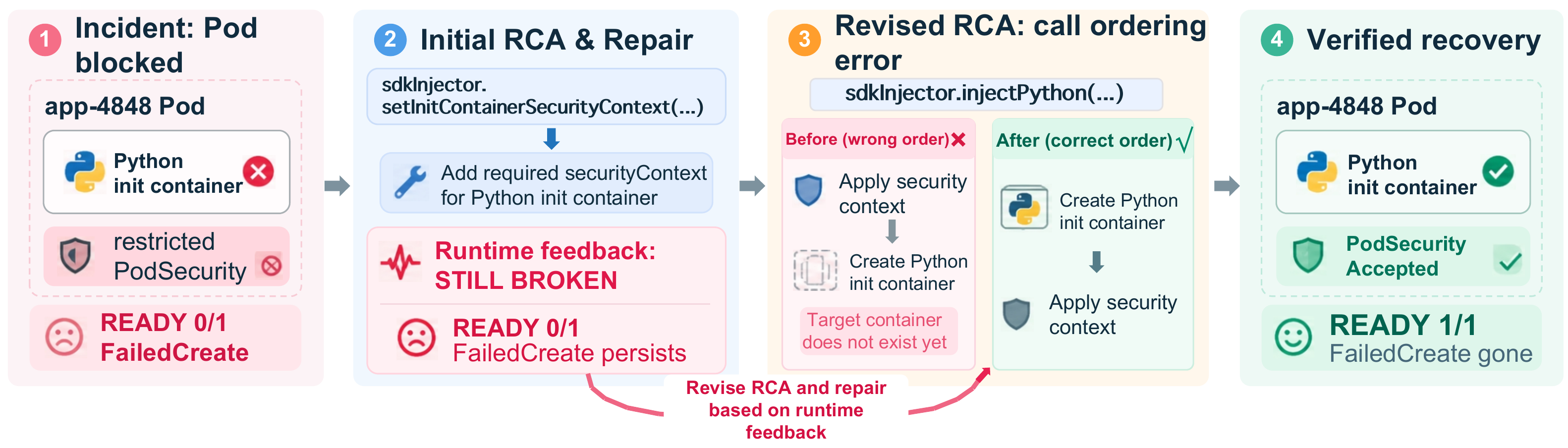}
    \vspace{-0.2in}
    \caption{Case A: An ineffective repair is corrected after the incident persists.
}
    \label{fig:case a}
    \vspace{-0.2in}
\end{figure*}
Figure~\ref{fig:rca_recovery}(a) reports cumulative system-side recovery across successive repair attempts, independent of whether the agent declares recovery. Claude Code + Claude-Sonnet-5 recovers 47.3\% of tasks within one repair attempt and 54.5\% within two, with no additional recoveries from later attempts. Codex + GPT-5.6 Luna recovers 36.4\% within one attempt, 40.0\% within two, and reaches 41.8\% within four. Thus, additional repair attempts can recover cases that remain unresolved after the initial intervention, although most observed gains occur early.

Figure~\ref{fig:rca_recovery}(b) complements this trajectory view with a task-level comparison of diagnosis and final Overall Recovery. For Codex + GPT-5.6 Luna, 14.5\% of tasks achieve recovery despite an incorrect joint RCA, whereas 20.0\% have a correct joint RCA but do not recover. For Claude Code + Claude-Sonnet-5, the corresponding fractions are 21.8\% and 7.3\%. Reported RCA correctness and final Overall Recovery therefore only partially coincide: an incorrect reported fault location can still be followed by a successful repair, while a correct diagnosis does not guarantee successful remediation.

\begin{figure*}[t]
    \centering
    \includegraphics[width=0.95\textwidth]{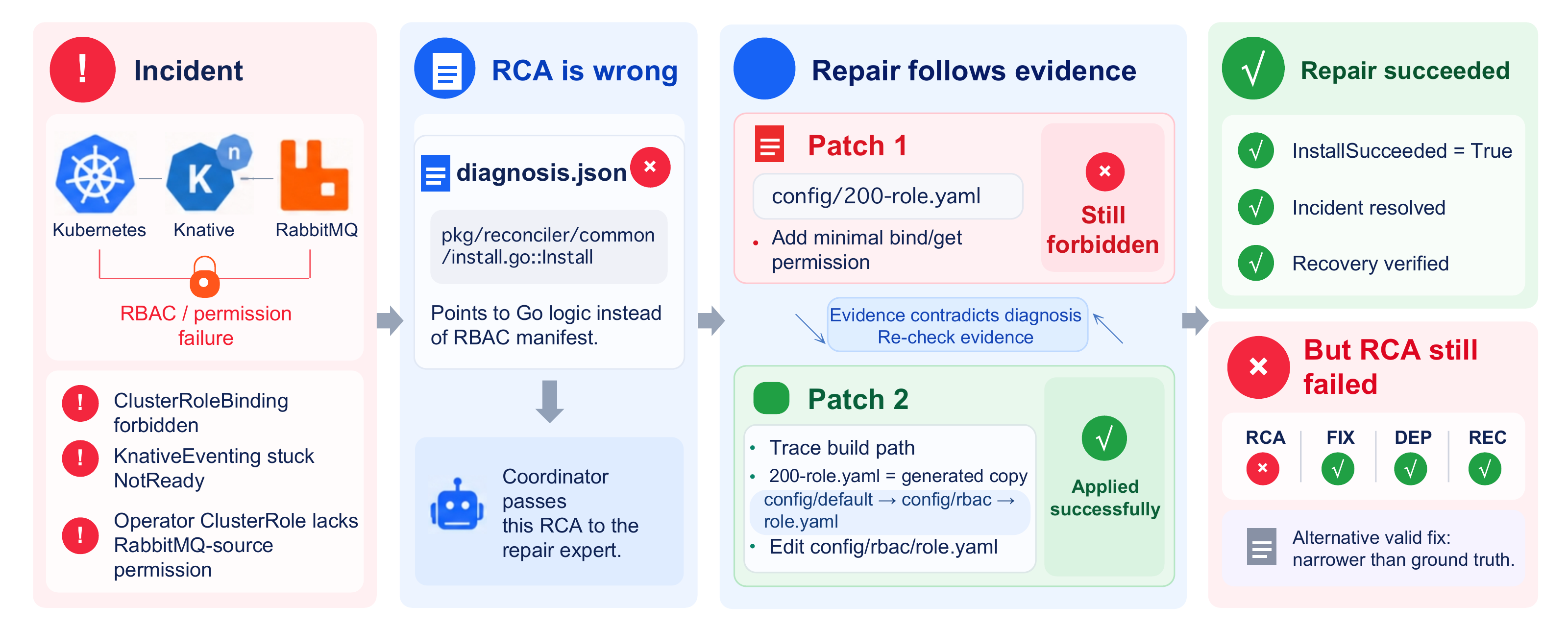}
    \vspace{-0.2in}
    \caption{Case B: Repair succeeds despite an incorrect RCA.}
    \label{fig:case-b}
    \vspace{-0.2in}
\end{figure*}

\begin{figure*}[t]
    \centering
    \includegraphics[width=0.9\textwidth]{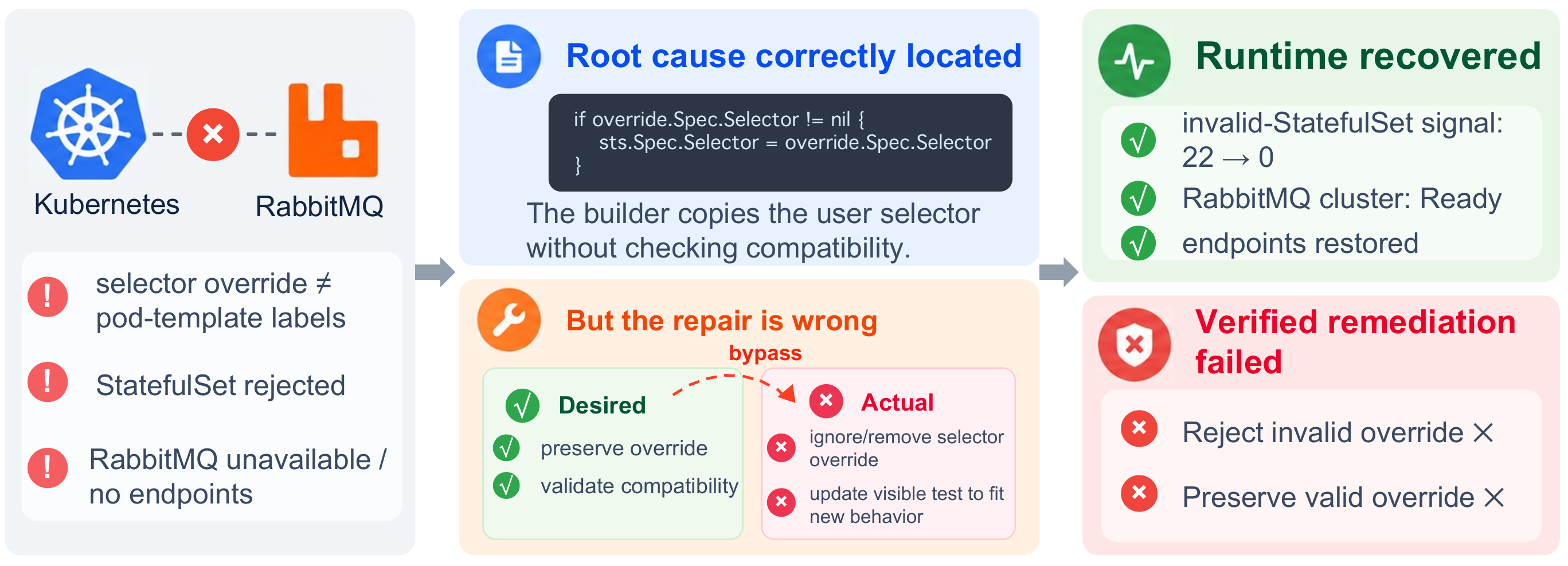}
    \vspace{-0.1in}
    \caption{Case C: Runtime symptoms disappear despite an incorrect repair.
}
    \label{fig:case c}
    \vspace{-0.3in}
\end{figure*}

The case studies illustrate how diagnosis, repair, and verification interact during execution. In Case A (Figure~\ref{fig:case a}), the initial repair remains ineffective after deployment, and the persistent failure leads the agent to revise its diagnosis and identify the ordering defect. In Case B (Figure~\ref{fig:case-b}), the agent eventually traces the authorization failure to the effective source manifest and repairs it, while its submitted RCA remains incorrect. Case C (Figure~\ref{fig:case c}) follows the opposite pattern: repair removes the immediate runtime symptom and restores the workload, but fix-specific verification later reveals that the intended behavior was not preserved. Together, these cases show that intermediate diagnostic and repair outcomes do not uniquely determine final resolution: runtime feedback can correct an ineffective intervention, successful remediation can emerge despite an incorrect reported RCA, and apparent recovery can still fail functional verification.

\section{Discussion}

\subsection{Implications}

\textbf{Maintain an explicit recovery contract.}
Our results suggest that incident-resolution agents should explicitly represent the behavior that a repair is expected to restore, rather than treating symptom disappearance as sufficient evidence of recovery. Such expectations can be derived from API definitions, configuration schemas, existing tests, and runtime observations, and should remain independent of candidate patches. This is particularly important for input-semantics defects, where removing an error may restore apparent availability while violating required behavior. An explicit recovery contract can therefore guide both repair generation and post-deployment verification.

\textbf{Treat diagnosis and repair as a feedback-driven loop.}
The case studies show that root-cause hypotheses should remain revisable throughout remediation. An initially plausible diagnosis may lead to an ineffective repair and be corrected through runtime feedback, while an incorrect reported RCA may still be followed by a successful repair. Future agents should therefore use deployment outcomes as evidence for updating their hypotheses rather than treating RCA as a one-time stage. They should also trace whether a source modification is realized in the built artifact, deployed revision, and expected runtime effect, helping distinguish an incorrect hypothesis from an intervention that was never fully realized.

\textbf{Target workflow mechanisms to observed failure modes.}
Different model--harness configurations exhibit different strengths, and substantially higher token consumption does not consistently yield better recovery. Together with frequent Action Realization Error and Insufficient Confirmation, this suggests that future workflows should target specific failure modes rather than uniformly adding more reasoning or tool use. For example, state tracking may reduce repeated investigation, while evidence-linked verification may reduce premature recovery declarations. Such mechanisms should be evaluated through matched-model and matched-budget comparisons.

\subsection{Potential Impact}

\textbf{Supporting integrated development and operations research.}
\nm provides a common experimental setting for studying agents that operate across both source code and live cloud systems. By coupling runtime observations, source-level repair, deployment feedback, and verified recovery, it enables researchers to study how diagnosis and remediation interact during execution rather than evaluating them as isolated capabilities. This setting can support joint investigation by software engineering and systems communities into agents that must reason across code and operational state.

\textbf{Turning historical defects into reusable research resources.}
Reproducing cloud incidents requires substantial effort to reconstruct software versions, dependencies, and triggering conditions. \nm packages confirmed historical defects into reusable executable environments that preserve their operational context and support controlled comparison of diagnosis and repair approaches. This reduces repeated reconstruction effort and provides a basis for extending evaluation to additional projects and newly confirmed defects.

\subsection{Threats to Validity}
\label{sec:threats-to-validity}

\textbf{Construct validity.}
Threats to construct validity concern whether our evaluation metrics accurately reflect incident resolution effectiveness and efficiency.
In \nm, task recovery requires passing code compilation, container deployment, fault oracle clearance, and test suites.
While these checks verify functional recovery under the triggering workload and guard against regressions, dynamic testing cannot mathematically prove behavioral correctness across every conceivable input condition.
A candidate patch might clear the fault probe while leaving an unexercised corner case unresolved.
Furthermore, the penalized latency metrics depend on predetermined stage weights ($B$, $3B$, and $6B$).
These metrics integrate milestone completion rates with elapsed time, serving as standardized evaluation scores rather than literal production outage durations.
Alternative penalty formulations could affect the relative efficiency rankings among agents.

\textbf{Internal validity.}
Threats to internal validity relate to experimental confounding factors, stochastic generation, and potential LLM training data contamination.
To reduce potential contamination from open-source repositories, \nm strictly withholds historical issue discussions and reference fixes from the execution environment.
Additionally, 27 of our 55 tasks originate from contemporary 2026 pull requests, narrowing the potential pre-training exposure window for the evaluated models, although this does not establish the absence of prior exposure.
The substantial computational overhead of repeatedly provisioning and executing live Kubernetes environments led us to evaluate one trial per task--configuration pair.
Consequently, the reported recovery rates and lifecycle latencies characterize the observed trajectories rather than expectations over repeated stochastic executions.
We therefore interpret differences among configurations descriptively rather than as statistically significant differences.
Repeated trials could quantify run-to-run variance and assess whether the observed configuration-level differences remain stable across stochastic executions.

\textbf{Conclusion validity.}
Threats to conclusion validity arise from the finite benchmark size and uneven task counts across incident categories.
Although aggregate metrics cover all 55 tasks, several subgroup analyses in Figure~\ref{fig:incident characteristics} rely on relatively small numbers of cases.
For example, Admission Policy contains three tasks, while Operator-reported and Storage blocked each contain five.
Consequently, percentage differences within these categories may correspond to only one or two tasks.
We therefore use the stratified results primarily to reveal heterogeneity across incident characteristics rather than to establish statistically stable differences between configurations.
Future extensions with additional incidents per category and repeated executions could support confidence intervals and more robust statistical comparisons.

\textbf{External validity.}
Threats to external validity concern the generalizability of our findings to other cloud environments and defect types.
Our benchmark focuses on confirmed software defects within 16 representative Kubernetes operator projects.
Consequently, our findings primarily characterize software-induced reconciliation failures in cloud management controllers, and may not directly generalize to infrastructure outages caused by hardware faults or to non-Kubernetes orchestration platforms.
Furthermore, our selection methodology excludes non-deterministic races and silent state drifts lacking observable runtime anomalies.

\section{Related Work}

Section~\ref{sec:prior_benchmarks_motivation} compares \nm with existing operational and software repair benchmarks.
This section contextualizes \nm within broader research on cloud incidents, automated failure diagnosis, and software repair.

\textbf{Cloud incidents.}
Empirical studies of production cloud systems show that incidents often stem from complex software bugs, configuration drift, and unanticipated runtime interactions rather than isolated hardware faults~\cite{incident2020,Incident2022SoCC,Microservie2018TSE}.
In cloud-native environments, the widespread adoption of softwarized management frameworks, such as Kubernetes operators, places complex reconciliation logic directly into operational loops~\cite{Sieve2022OSDI,Acto2023SOSP,OAT2026NSDI,Kivi2024ATC}.
Defects in these controllers lead to runtime failures when reconciling desired and observed system state under dynamic workloads.
Operational actions can restore service without correcting the defective implementation, making service recovery and implementation repair distinct concerns.
\nm connects these concerns by reproducing confirmed historical operator defects and evaluating source or configuration repairs through their effects on the running system.

\textbf{Automated failure diagnosis.}
Automated failure diagnosis uses runtime evidence to detect anomalies and identify their causes.
Related methods and benchmarks examine metrics, logs, and distributed traces as sources of diagnostic information~\cite{nezha2023fse,OpenRCA2025ICLR,RCA2025Eval}.
Telemetry-based techniques exploit service dependencies, execution traces, compact data representations, and learned failure patterns to localize faults~\cite{yu2021microrank,Dejavu2022FSE,MicroSketch2022}.
LLM-based approaches combine diagnostic information collection with explanatory reasoning and tool-assisted investigation~\cite{RCACopilot,RCAgent2024CIKM,COCA}.
Further work introduces procedural guidance, structured representations of Kubernetes state, and multi-agent fault diagnosis~\cite{FlowofAction2025WWW,SynergyRCA,LLMGuard}.
These approaches improve the ability to interpret operational evidence and produce diagnostic conclusions.
Assessing a diagnosis, however, does not by itself establish that a subsequent repair restores the affected functionality.
\nm evaluates this connection through patch generation, deployment, and recovery verification within the same incident-resolution process.

\textbf{Software repair agents.}
Automated program repair investigates techniques to eliminate software defects with minimal human intervention.
Search-based and constraint-based approaches use tests to guide candidate generation through genetic programming or semantic analysis~\cite{GenProg2012TSE,SemFix2013ICSE}.
Learning-based repair incorporates project-specific knowledge and execution diagnostics, while conversational approaches use compilation and test feedback to revise unsuccessful candidates~\cite{SelfAPR2022ASE,ChatRepair2024ISSTA}.
Historical defect datasets and issue-resolution benchmarks provide reproducible subjects and executable tests for repair evaluation~\cite{Defects4J,SWEBench2024ICLR}.
Repository-level methods combine code search, contextual investigation, patch generation, and execution feedback through interactive or structured workflows~\cite{AutoCodeRover2024ISSTA,SWEAgent,Agentless,RepairAgent2025ICSE}.
These methods already establish iterative repair and execution feedback as important capabilities.
\nm evaluates these capabilities within an incident-resolution lifecycle that begins with observations from a running cluster.
Candidate changes must be built and deployed, and their effects are assessed through runtime fault clearance, regression testing, and workload recovery.

\vspace{-0.2in}

\section{Conclusion}

This paper introduces \nm, an interactive benchmark for evaluating autonomous agents across the cloud incident resolution lifecycle.
As autonomous agents advance, the boundary between cloud operations and software development diminishes, demanding integrated failure detection, localization, and source-level repair.
Grounded in \num reproducible historical defects from 16 Kubernetes operator projects, \nm combines interactive cluster environments with executable oracles covering runtime fault clearance and functional regression suites.
Across six evaluated agent–model configurations, overall recovery ranges from 14.5\% to 52.7\%, showing that successful incident recognition and source-level diagnosis do not reliably translate into end-to-end resolution. Trajectory analysis further demonstrates that reported diagnosis and remediation are only partially coupled, and that runtime symptom clearance does not necessarily preserve the required functionality. 
By releasing the benchmark artifacts 
and evaluation infrastructure, \nm provides a foundation for developing more reliable agents in integrated development and operational settings.

\section{Data Availability}
The \nm benchmark and accompanying experimental artifacts are available in our repository at https://github.com/LLM4Ops/SWEOps-Bench.


\bibliographystyle{ACM-Reference-Format}
\bibliography{ref}

@inproceedings{yu2021microrank,
  author = {Yu, Guangba and Chen, Pengfei and Chen, Hongyang and Guan, Zijie and Huang, Zicheng and Jing, Linxiao and Weng, Tianjun and Sun, Xinmeng and Li, Xiaoyun},
  title = {MicroRank: End-to-End Latency Issue Localization with Extended Spectrum Analysis in Microservice Environments},
  year = {2021},
  publisher = {ACM},
  booktitle = {WWW 2021},
  pages = {3087–3098},
  numpages = {12},
  doi      = {10.1145/3442381.3449905},
}

@inproceedings{nezha2023fse,
  author       = {Guangba Yu and
                  Pengfei Chen and
                  Yufeng Li and
                  Hongyang Chen and
                  Xiaoyun Li and
                  Zibin Zheng},
  title        = {Nezha: Interpretable Fine-Grained Root Causes Analysis for Microservices on Multi-modal Observability Data},
  booktitle    = {{ESEC/FSE} 2023},
  pages        = {553--565},
  publisher    = {{ACM}},
  year         = {2023},
  url          = {https://doi.org/10.1145/3611643.3616249},
  doi          = {10.1145/3611643.3616249},
}

@inproceedings{li2022incident,
  title={Going through the Life Cycle of Faults in Clouds: Guidelines on Fault Handling},
  author={Li, Xiaoyun and Yu, Guangba and Chen, Pengfei and Chen, Hongyang and Chen, Zhekang},
  booktitle={ISSRE 2022},
  pages={121--132},
  year={2022},
  publisher={IEEE},
  doi          = {10.1109/ISSRE55969.2022.00022},
  url          = {https://doi.org/10.1109/ISSRE55969.2022.00022},
}

@inproceedings{OpenRCA2025ICLR,
  author       = {Junjielong Xu and
                  Qinan Zhang and
                  Zhiqing Zhong and
                  Shilin He and
                  Chaoyun Zhang and
                  Qingwei Lin and
                  Dan Pei and
                  Pinjia He and
                  Dongmei Zhang and
                  Qi Zhang},
  title        = {OpenRCA: Can Large Language Models Locate the Root Cause of Software Failures?},
  booktitle    = {{ICLR} 2025},
  publisher    = {OpenReview.net},
  year         = {2025},
  url          = {https://openreview.net/forum?id=M4qNIzQYpd},
  bibsource    = {dblp computer science bibliography, https://dblp.org}
}

@inproceedings{chen2025aiopslab,
title={{AIO}psLab: A Holistic Framework to Evaluate {AI} Agents for Enabling Autonomous Clouds},
author={Yinfang Chen and Manish Shetty and Gagan Somashekar and Minghua Ma and Yogesh Simmhan and Jonathan Mace and Chetan Bansal and Rujia Wang and Saravan Rajmohan},
booktitle={{MLSys} 2025},
year={2025},
url={https://openreview.net/forum?id=3EXBLwGxtq}
}

@inproceedings{ITBench2025ICML,
  author       = {Saurabh Jha and
                  Rohan R. Arora and
                  Yuji Watanabe and
                  Takumi Yanagawa and
                  Yinfang Chen and
                  Jackson Clark and
                  Bhavya and
                  Mudit Verma and
                  Harshit Kumar and
                  Hirokuni Kitahara and
                  Noah Zheutlin and
                  Saki Takano and
                  Divya Pathak and
                  Felix George and
                  Xinbo Wu and
                  Bekir O. Turkkan and
                  Gerard Vanloo and
                  Michael Nidd and
                  Ting Dai and
                  Oishik Chatterjee and
                  Pranjal Gupta and
                  Suranjana Samanta and
                  Pooja Aggarwal and
                  Rong Lee and
                  Jae{-}wook Ahn and
                  Debanjana Kar and
                  Amit M. Paradkar and
                  Yu Deng and
                  Pratibha Moogi and
                  Prateeti Mohapatra and
                  Naoki Abe and
                  Chandrasekhar Narayanaswami and
                  Tianyin Xu and
                  Lav R. Varshney and
                  Ruchi Mahindru and
                  Anca Sailer and
                  Laura Shwartz and
                  Daby Sow and
                  Nicholas C. Fuller and
                  Ruchir Puri},
  title        = {ITBench: Evaluating {AI} Agents across Diverse Real-World {IT} Automation
                  Tasks},
  booktitle    = {{ICML} 2025},
  publisher    = {OpenReview.net},
  year         = {2025},
  url          = {https://openreview.net/forum?id=jP59rz1bZk},
}

@inproceedings{OpsEval2025FSE,
  author       = {Yuhe Liu and
                  Changhua Pei and
                  Longlong Xu and
                  Bohan Chen and
                  Mingze Sun and
                  Zhirui Zhang and
                  Yongqian Sun and
                  Shenglin Zhang and
                  Kun Wang and
                  Haiming Zhang and
                  Jianhui Li and
                  Gaogang Xie and
                  Xidao Wen and
                  Xiaohui Nie and
                  Minghua Ma and
                  Dan Pei},
  title        = {OpsEval: {A} Comprehensive Benchmark Suite for Evaluating Large Language
                  Models' Capability in {IT} Operations Domain},
  booktitle    = { {FSE} Companion 2025},
  pages        = {503--513},
  publisher    = {{ACM}},
  year         = {2025},
  url          = {https://doi.org/10.1145/3696630.3728572},
  doi          = {10.1145/3696630.3728572},
}

@inproceedings{RCA2025Eval,
  author       = {Luan Pham and
                  Hongyu Zhang and
                  Huong Ha and
                  Flora Salim and
                  Xiuzhen Zhang},
  title        = {RCAEval: {A} Benchmark for Root Cause Analysis of Microservice Systems
                  with Telemetry Data},
  booktitle    = {{WWW} 2025},
  pages        = {777--780},
  publisher    = {{ACM}},
  year         = {2025},
  url          = {https://doi.org/10.1145/3701716.3715290},
  doi          = {10.1145/3701716.3715290},
}

@inproceedings{MicroSketch2022,
author = {Li, Yufeng and Yu, Guangba and Chen, Pengfei and Zhang, Chuanfu and Zheng, Zibin},
title = {MicroSketch: Lightweight and Adaptive Sketch Based Performance Issue Detection Localization in Microservice Systems},
year = {2022},
isbn = {978-3-031-20983-3},
publisher = {Springer-Verlag},
address = {Berlin, Heidelberg},
url = {https://doi.org/10.1007/978-3-031-20984-0_15},
doi = {10.1007/978-3-031-20984-0_15},
booktitle = {ICSOC 2022},
pages = {219–236},
}

@inproceedings{COCA,
  author       = {Yichen Li and
                  Yulun Wu and
                  Jinyang Liu and
                  Zhihan Jiang and
                  Zhuangbin Chen and
                  Guangba Yu and
                  Michael R. Lyu},
  title        = {{COCA:} Generative Root Cause Analysis for Distributed Systems with
                  Code Knowledge},
  booktitle    = {{ICSE} 2025},
  pages        = {1346--1358},
  publisher    = {{IEEE}},
  year         = {2025},
  url          = {https://doi.org/10.1109/ICSE55347.2025.00234},
  doi          = {10.1109/ICSE55347.2025.00234},
}

@inproceedings{RCACopilot,
  author       = {Yinfang Chen and
                  Huaibing Xie and
                  Minghua Ma and
                  Yu Kang and
                  Xin Gao and
                  Liu Shi and
                  Yunjie Cao and
                  Xuedong Gao and
                  Hao Fan and
                  Ming Wen and
                  Jun Zeng and
                  Supriyo Ghosh and
                  Xuchao Zhang and
                  Chaoyun Zhang and
                  Qingwei Lin and
                  Saravan Rajmohan and
                  Dongmei Zhang and
                  Tianyin Xu},
  title        = {Automatic Root Cause Analysis via Large Language Models for Cloud
                  Incidents},
  booktitle    = {{EuroSys} 2024},
  pages        = {674--688},
  publisher    = {{ACM}},
  year         = {2024},
  url          = {https://doi.org/10.1145/3627703.3629553},
  doi          = {10.1145/3627703.3629553},
}

@inproceedings{FlowofAction2025WWW,
author = {Pei, Changhua and Wang, Zexin and Liu, Fengrui and Li, Zeyan and Liu, Yang and He, Xiao and Kang, Rong and Zhang, Tieying and Chen, Jianjun and Li, Jianhui and Xie, Gaogang and Pei, Dan},
title = {Flow-of-Action: SOP Enhanced LLM-Based Multi-Agent System for Root Cause Analysis},
year = {2025},
publisher = {Association for Computing Machinery},
url = {https://doi.org/10.1145/3701716.3715225},
doi = {10.1145/3701716.3715225},
booktitle = {{WWW} 2025},
pages = {422–431},
}

@misc{alertguardian2025ASE,
      title={AlertGuardian: Intelligent Alert Life-Cycle Management for Large-scale Cloud Systems}, 
      author={Guangba Yu and Genting Mai and Rui Wang and Ruipeng Li and Pengfei Chen and Long Pan and Ruijie Xu},
      year={2026},
      eprint={2601.14912},
      archivePrefix={arXiv},
      primaryClass={cs.DC},
      url={https://arxiv.org/abs/2601.14912}, 
}

@inproceedings{RCAgent2024CIKM,
  author       = {Zefan Wang and
                  Zichuan Liu and
                  Yingying Zhang and
                  Aoxiao Zhong and
                  Jihong Wang and
                  Fengbin Yin and
                  Lunting Fan and
                  Lingfei Wu and
                  Qingsong Wen},
  title        = {RCAgent: Cloud Root Cause Analysis by Autonomous Agents with Tool-Augmented Large Language Models},
  booktitle    = {{CIKM} 2024},
  pages        = {4966--4974},
  publisher    = {{ACM}},
  year         = {2024},
  url          = {https://doi.org/10.1145/3627673.3680016},
  doi          = {10.1145/3627673.3680016},
}

@article{SynergyRCA,
  author       = {Yong Xiang and
                  Charley Peter Chen and
                  Liyi Zeng and
                  Wei Yin and
                  Xin Liu and
                  Hu Li and
                  Wei Xu},
  title        = {Simplifying Root Cause Analysis in Kubernetes with StateGraph and {LLM}},
  journal      = {CoRR},
  volume       = {abs/2506.02490},
  year         = {2025},
  url          = {https://doi.org/10.48550/arXiv.2506.02490},
  doi          = {10.48550/ARXIV.2506.02490},
  eprinttype    = {arXiv},
  eprint       = {2506.02490},
}

@inproceedings{incident2020,
  author       = {Junjie Chen and
                  Shu Zhang and
                  Xiaoting He and
                  Qingwei Lin and
                  Hongyu Zhang and
                  Dan Hao and
                  Yu Kang and
                  Feng Gao and
                  Zhangwei Xu and
                  Yingnong Dang and
                  Dongmei Zhang},
  title        = {How Incidental are the Incidents? Characterizing and Prioritizing Incidents for Large-Scale Online Service Systems},
  booktitle    = {{ASE} 2020},
  pages        = {373--384},
  publisher    = {{IEEE}},
  year         = {2020},
  url          = {https://doi.org/10.1145/3324884.3416624},
  doi          = {10.1145/3324884.3416624},
}

@inproceedings{Dejavu2022FSE,
  author       = {Zeyan Li and
                  Nengwen Zhao and
                  Mingjie Li and
                  Xianglin Lu and
                  Lixin Wang and
                  Dongdong Chang and
                  Xiaohui Nie and
                  Li Cao and
                  Wenchi Zhang and
                  Kaixin Sui and
                  Yanhua Wang and
                  Xu Du and
                  Guoqiang Duan and
                  Dan Pei},
  title        = {Actionable and interpretable fault localization for recurring failures in online service systems},
  booktitle    = {{ESEC/FSE} 2022},
  pages        = {996--1008},
  publisher    = {{ACM}},
  year         = {2022},
  url          = {https://doi.org/10.1145/3540250.3549092},
  doi          = {10.1145/3540250.3549092},

}

@article{Microservie2018TSE,
  author       = {Xiang Zhou and
                  Xin Peng and
                  Tao Xie and
                  Jun Sun and
                  Chao Ji and
                  Wenhai Li and
                  Dan Ding},
  title        = {Fault Analysis and Debugging of Microservice Systems: Industrial Survey,  Benchmark System, and Empirical Study},
  journal      = {{IEEE} Trans. Software Eng.},
  volume       = {47},
  number       = {2},
  pages        = {243--260},
  year         = {2021},
  url          = {https://doi.org/10.1109/TSE.2018.2887384},
  doi          = {10.1109/TSE.2018.2887384},
  bibsource    = {dblp computer science bibliography, https://dblp.org}
}

@INPROCEEDINGS{LLMGuard,
  author={Zhong, Yuedong and Yu, Guangba and Huang, Yujie and Fu, QunChao and Ren, Rui and Feng, Cong and Yang, Yongqiang and Lyu, Michael},
  booktitle={2026 56th Annual IEEE International Conference on Dependable Systems and Networks (DSN)}, 
  title={LLMGuard: Multi-Agent Fault Diagnosis for Reliable Language-Model-as-a-Service}, 
  year={2026},
  volume={},
  number={},
  pages={15-23},
  doi={10.1109/DSN69566.2026.00018}}

@misc{redisOperator1540,
  author       = {{Redis-Operator}},
  title        = {{RedisCluster} Pods Aren't Created with No Persistence and Using {initContainer}},
  year         = {2025},
  howpublished = {GitHub issue \#1540},
  url          = {https://github.com/OT-CONTAINER-KIT/redis-operator/issues/1540}
}

@inproceedings {Sieve2022OSDI,
author = {Xudong Sun and Wenqing Luo and Jiawei Tyler Gu and Aishwarya Ganesan and Ramnatthan Alagappan and Michael Gasch and Lalith Suresh and Tianyin Xu},
title = {Automatic Reliability Testing For Cluster Management Controllers},
booktitle = {16th USENIX Symposium on Operating Systems Design and Implementation (OSDI 22)},
year = {2022},
isbn = {978-1-939133-28-1},
address = {Carlsbad, CA},
pages = {143--159},
url = {https://www.usenix.org/conference/osdi22/presentation/sun},
publisher = {USENIX Association},
}

@inproceedings{Acto2023SOSP,
author = {Gu, Jiawei Tyler and Sun, Xudong and Zhang, Wentao and Jiang, Yuxuan and Wang, Chen and Vaziri, Mandana and Legunsen, Owolabi and Xu, Tianyin},
title = {Acto: Automatic End-to-End Testing for Operation Correctness of Cloud System Management},
year = {2023},
isbn = {9798400702297},
publisher = {Association for Computing Machinery},
address = {New York, NY, USA},
url = {https://doi.org/10.1145/3600006.3613161},
doi = {10.1145/3600006.3613161},
booktitle = {Proceedings of the 29th Symposium on Operating Systems Principles},
pages = {96–112},
numpages = {17},
location = {Koblenz, Germany},
series = {SOSP '23}
}

@inproceedings{OAT2026NSDI,
  author       = {Jiawei Tyler Gu and
                  Zhen Tang and
                  Yiming Su and
                  Bogdan Alexandru Stoica and
                  Xudong Sun and
                  William X. Zheng and
                  Yue Zhang and
                  Akond Rahman and
                  Chen Wang and
                  Tianyin Xu},
  editor       = {Srikanth Kandula and
                  Hakim Weatherspoon},
  title        = {Who Watches the Watchers? On the Reliability of Softwarizing Cloud
                  Application Management},
  booktitle    = {23rd {USENIX} Symposium on Networked Systems Design and Implementation,
                  {NSDI} 2026, Renton, WA, May 4-6, 2026},
  pages        = {1171--1190},
  publisher    = {{USENIX} Association},
  year         = {2026},
  url          = {https://www.usenix.org/conference/nsdi26/presentation/gu},
  bibsource    = {dblp computer science bibliography, https://dblp.org}
}

@inproceedings{Kivi2024ATC,
author = {Liu, Bingzhe and Lim, Gangmuk and Beckett, Ryan and Godfrey, P. Brighten},
title = {Kivi: verification for cluster management},
year = {2024},
isbn = {978-1-939133-41-0},
publisher = {USENIX Association},
address = {USA},
booktitle = {Proceedings of the 2024 USENIX Conference on Usenix Annual Technical Conference},
articleno = {32},
numpages = {19},
location = {Santa Clara, CA, USA},
series = {USENIX ATC'24}
}

@misc{AtlassianIncidentResponse,
  author       = {{Atlassian}},
  title        = {Incident Response: How We Respond to an Incident},
  howpublished = {Atlassian Incident Management Handbook},
  year         = {2024},
  url          = {https://www.atlassian.com/incident-management/handbook/incident-response}
}

@inproceedings{Incident2022SoCC,
author = {Ghosh, Supriyo and Shetty, Manish and Bansal, Chetan and Nath, Suman},
title = {How to fight production incidents? an empirical study on a large-scale cloud service},
year = {2022},
isbn = {9781450394147},
publisher = {Association for Computing Machinery},
address = {New York, NY, USA},
url = {https://doi.org/10.1145/3542929.3563482},
doi = {10.1145/3542929.3563482},
booktitle = {Proceedings of the 13th Symposium on Cloud Computing},
pages = {126–141},
numpages = {16},
location = {San Francisco, California},
series = {SoCC '22}
}

@misc{LogicMonitorMTTR,
  author       = {{LogicMonitor}},
  title        = {What's the Difference Between {MTTR}, {MTBF}, {MTTD}, and {MTTF}},
  howpublished = {LogicMonitor Blog},
  year         = {2024},
  url          = {https://www.logicmonitor.com/blog/whats-the-difference-between-mttr-mttd-mttf-and-mtbf}
}

@article{Avizienis2004Taxonomy,
author = {Avizienis, Algirdas and Laprie, Jean-Claude and Randell, Brian and Landwehr, Carl},
title = {Basic Concepts and Taxonomy of Dependable and Secure Computing},
year = {2004},
issue_date = {January 2004},
publisher = {IEEE Computer Society Press},
address = {Washington, DC, USA},
volume = {1},
number = {1},
issn = {1545-5971},
url = {https://doi.org/10.1109/TDSC.2004.2},
doi = {10.1109/TDSC.2004.2},
journal = {IEEE Trans. Dependable Secur. Comput.},
month = jan,
pages = {11–33},
numpages = {23}
}

@book{ITILIncidentManagement,
  author    = {{AXELOS}},
  title     = {ITIL Foundation: ITIL 4 Edition},
  publisher = {TSO (The Stationery Office)},
  year      = {2019}
}

@misc{wang2026cloudopsbench,
      title={Cloud-OpsBench: A Reproducible Benchmark for Agentic Root Cause Analysis in Cloud Systems}, 
      author={Yilun Wang and Guangba Yu and Haiyu Huang and Yujie Huang and Zirui Wang and Pengfei Chen and Michael R. Lyu},
      year={2026},
      eprint={2603.00468},
      archivePrefix={arXiv},
      primaryClass={cs.SE},
      url={https://arxiv.org/abs/2603.00468}, 
}

@inproceedings{SWEBench2024ICLR,
  author       = {Carlos E. Jimenez and
                  John Yang and
                  Alexander Wettig and
                  Shunyu Yao and
                  Kexin Pei and
                  Ofir Press and
                  Karthik R. Narasimhan},
  title        = {SWE-bench: Can Language Models Resolve Real-world Github Issues?},
  booktitle    = {The Twelfth International Conference on Learning Representations,
                  {ICLR} 2024, Vienna, Austria, May 7-11, 2024},
  publisher    = {OpenReview.net},
  year         = {2024},
  url          = {https://openreview.net/forum?id=VTF8yNQM66},
  bibsource    = {dblp computer science bibliography, https://dblp.org}
}

@inproceedings{yang2025swebench,
    title={SWE-bench Multimodal: Do AI Systems Generalize to Visual Software Domains?},
    author={John Yang and Carlos E Jimenez and Alex L Zhang and Kilian Lieret and Joyce Yang and Xindi Wu and Ori Press and Niklas Muennighoff and Gabriel Synnaeve and Karthik R Narasimhan and Diyi Yang and Sida Wang and Ofir Press},
    booktitle={The Thirteenth International Conference on Learning Representations},
    year={2025},
    url={https://openreview.net/forum?id=riTiq3i21b}
}

@article{PAR,
  author       = {Frank Hutter and
                  Holger H. Hoos and
                  Kevin Leyton{-}Brown and
                  Thomas St{\"{u}}tzle},
  title        = {ParamILS: An Automatic Algorithm Configuration Framework},
  journal      = {J. Artif. Intell. Res.},
  volume       = {36},
  pages        = {267--306},
  year         = {2009},
  url          = {https://doi.org/10.1613/jair.2861},
  doi          = {10.1613/JAIR.2861},
  bibsource    = {dblp computer science bibliography, https://dblp.org}
}

@software{Harbor,
author = {{Harbor Framework Team}},
title = {{Harbor: A framework for evaluating and optimizing agents and models in container environments}},
year = {2026},
doi = {10.5281/zenodo.20953922},
url = {https://doi.org/10.5281/zenodo.20953922}
}

@article{GenProg2012TSE,
author = {Le Goues, Claire and Nguyen, ThanhVu and Forrest, Stephanie and Weimer, Westley},
title = {GenProg: A Generic Method for Automatic Software Repair},
year = {2012},
issue_date = {January 2012},
publisher = {IEEE Press},
volume = {38},
number = {1},
issn = {0098-5589},
url = {https://doi.org/10.1109/TSE.2011.104},
doi = {10.1109/TSE.2011.104},
journal = {IEEE Trans. Softw. Eng.},
month = jan,
pages = {54–72},
numpages = {19}
}

@inproceedings{SemFix2013ICSE,
author = {Nguyen, Hoang Duong Thien and Qi, Dawei and Roychoudhury, Abhik and Chandra, Satish},
title = {SemFix: program repair via semantic analysis},
year = {2013},
isbn = {9781467330763},
publisher = {IEEE Press},
booktitle = {Proceedings of the 2013 International Conference on Software Engineering},
pages = {772–781},
numpages = {10},
location = {San Francisco, CA, USA},
series = {ICSE '13}
}

@inproceedings{SelfAPR2022ASE,
  author       = {He Ye and
                  Matias Martinez and
                  Xiapu Luo and
                  Tao Zhang and
                  Martin Monperrus},
  title        = {SelfAPR: Self-supervised Program Repair with Test Execution Diagnostics},
  booktitle    = {37th {IEEE/ACM} International Conference on Automated Software Engineering,
                  {ASE} 2022, Rochester, MI, USA, October 10-14, 2022},
  pages        = {92:1--92:13},
  publisher    = {{ACM}},
  year         = {2022},
  url          = {https://doi.org/10.1145/3551349.3556926},
  doi          = {10.1145/3551349.3556926},
  bibsource    = {dblp computer science bibliography, https://dblp.org}
}

@inproceedings{ChatRepair2024ISSTA,
  author       = {Chunqiu Steven Xia and
                  Lingming Zhang},
  editor       = {Maria Christakis and
                  Michael Pradel},
  title        = {Automated Program Repair via Conversation: Fixing 162 out of 337 Bugs
                  for {\textdollar}0.42 Each using ChatGPT},
  booktitle    = {Proceedings of the 33rd {ACM} {SIGSOFT} International Symposium on
                  Software Testing and Analysis, {ISSTA} 2024, Vienna, Austria, September
                  16-20, 2024},
  pages        = {819--831},
  publisher    = {{ACM}},
  year         = {2024},
  url          = {https://doi.org/10.1145/3650212.3680323},
  doi          = {10.1145/3650212.3680323},
  bibsource    = {dblp computer science bibliography, https://dblp.org}
}

@inproceedings{Defects4J,
  author       = {Ren{\'{e}} Just and
                  Darioush Jalali and
                  Michael D. Ernst},
  editor       = {Corina S. Pasareanu and
                  Darko Marinov},
  title        = {Defects4J: a database of existing faults to enable controlled testing
                  studies for Java programs},
  booktitle    = {International Symposium on Software Testing and Analysis, {ISSTA}
                  '14, San Jose, CA, {USA} - July 21 - 26, 2014},
  pages        = {437--440},
  publisher    = {{ACM}},
  year         = {2014},
  url          = {https://doi.org/10.1145/2610384.2628055},
  doi          = {10.1145/2610384.2628055},
  bibsource    = {dblp computer science bibliography, https://dblp.org}
}

@inproceedings{AutoCodeRover2024ISSTA,
  author       = {Yuntong Zhang and
                  Haifeng Ruan and
                  Zhiyu Fan and
                  Abhik Roychoudhury},
  editor       = {Maria Christakis and
                  Michael Pradel},
  title        = {AutoCodeRover: Autonomous Program Improvement},
  booktitle    = {Proceedings of the 33rd {ACM} {SIGSOFT} International Symposium on
                  Software Testing and Analysis, {ISSTA} 2024, Vienna, Austria, September
                  16-20, 2024},
  pages        = {1592--1604},
  publisher    = {{ACM}},
  year         = {2024},
  url          = {https://doi.org/10.1145/3650212.3680384},
  doi          = {10.1145/3650212.3680384},
  bibsource    = {dblp computer science bibliography, https://dblp.org}
}

@inproceedings{SWEAgent,
 author = {Yang, John and Jimenez, Carlos and Wettig, Alexander and Lieret, Kilian and Yao, Shunyu and Narasimhan, Karthik and Press, Ofir},
 booktitle = {Advances in Neural Information Processing Systems},
 doi = {10.52202/079017-1601},
 editor = {A. Globerson and L. Mackey and D. Belgrave and A. Fan and U. Paquet and J. Tomczak and C. Zhang},
 pages = {50528--50652},
 publisher = {Curran Associates, Inc.},
 title = {SWE-agent: Agent-Computer Interfaces Enable Automated Software Engineering},
 url = {https://proceedings.neurips.cc/paper_files/paper/2024/file/5a7c947568c1b1328ccc5230172e1e7c-Paper-Conference.pdf},
 volume = {37},
 year = {2024}
}

@article{Agentless,
  author       = {Chunqiu Steven Xia and
                  Yinlin Deng and
                  Soren Dunn and
                  Lingming Zhang},
  title        = {Demystifying LLM-Based Software Engineering Agents},
  journal      = {Proc. {ACM} Softw. Eng.},
  volume       = {2},
  number       = {{FSE}},
  pages        = {801--824},
  year         = {2025},
  url          = {https://doi.org/10.1145/3715754},
  doi          = {10.1145/3715754},
  bibsource    = {dblp computer science bibliography, https://dblp.org}
}

@inproceedings{RepairAgent2025ICSE,
  author       = {Islem Bouzenia and
                  Premkumar T. Devanbu and
                  Michael Pradel},
  title        = {RepairAgent: An Autonomous, LLM-Based Agent for Program Repair},
  booktitle    = {47th {IEEE/ACM} International Conference on Software Engineering,
                  {ICSE} 2025, Ottawa, ON, Canada, April 26 - May 6, 2025},
  pages        = {2188--2200},
  publisher    = {{IEEE}},
  year         = {2025},
  url          = {https://doi.org/10.1109/ICSE55347.2025.00157},
  doi          = {10.1109/ICSE55347.2025.00157},
  bibsource    = {dblp computer science bibliography, https://dblp.org}
}

@misc{redisOperator1864,
  author       = {{OT-CONTAINER-KIT}},
  title        = {fix(utils): pass client certificate to exec-based redis-cli},
  year         = {2026},
  howpublished = {GitHub pull request \#1864},
  url          = {https://github.com/OT-CONTAINER-KIT/redis-operator/pull/1864}
}

@inproceedings{terminalbench,
  author       = {Mike A. Merrill and
                  Alexander Glenn Shaw and
                  Nicholas Carlini and
                  Boxuan Li and
                  Harsh Raj and
                  Ivan Bercovich and
                  Lin Shi and
                  Jeong Yeon Shin and
                  Thomas Walshe and
                  Estefany Kelly Buchanan and
                  Junhong Shen and
                  Guanghao Ye and
                  Haowei Lin and
                  Jason Poulos and
                  Maoyu Wang and
                  Marianna Nezhurina and
                  Di Lu and
                  Orfeas Menis{-}Mastromichalakis and
                  Zhiwei Xu and
                  Zizhao Chen and
                  Yue Liu and
                  Robert Zhang and
                  Leon Liangyu Chen and
                  Anurag Kashyap and
                  Jan{-}Lucas Uslu and
                  Jeffrey Li and
                  Jianbo Wu and
                  Minghao Yan and
                  Song Bian and
                  Vedang Sharma and
                  Ke Sun and
                  Steven Dillmann and
                  Akshay Anand and
                  Andrew Lanpouthakoun and
                  Bardia Koopah and
                  Changran Hu and
                  Etash Kumar Guha and
                  Gabriel H. S. Dreiman and
                  Jiacheng Zhu and
                  Karl Krauth and
                  Li Zhong and
                  Niklas Muennighoff and
                  Robert Amanfu and
                  Shangyin Tan and
                  Shreyas Pimpalgaonkar and
                  Tushar Aggarwal and
                  Xiangning Lin and
                  Xin Lan and
                  Xuandong Zhao and
                  Yiqing Liang and
                  Yuanli Wang and
                  Zilong Wang and
                  Changzhi Zhou and
                  David Heineman and
                  Hange Liu and
                  Harsh Trivedi and
                  John Yang and
                  Junhong Lin and
                  Manish Shetty and
                  Michael Yang and
                  Nabil Omi and
                  Negin Raoof and
                  Shanda Li and
                  Terry Yue Zhuo and
                  Wuwei Lin and
                  Yiwei Dai and
                  Yuxin Wang and
                  Wenhao Chai and
                  Shang Zhou and
                  Dariush Wahdany and
                  Ziyu She and
                  Jiaming Hu and
                  Zhikang Dong and
                  Yuxuan Zhu and
                  Sasha Cui and
                  Ahson Saiyed and
                  Arinbj{\"{o}}rn Kolbeinsson and
                  Christopher Michael Rytting and
                  Ryan Marten and
                  Yixin Wang and
                  Jenia Jitsev and
                  Alex Dimakis and
                  Andy Konwinski and
                  Ludwig Schmidt},
  editor       = {Carl Vondrick and
                  Bharath Hariharan and
                  Colin Raffel and
                  Lerrel Pinto and
                  Diyi Yang and
                  Aleksandra Faust},
  title        = {Terminal-Bench: Benchmarking Agents on Hard, Realistic Tasks in Command
                  Line Interfaces},
  booktitle    = {The Fourteenth International Conference on Learning Representations,
                  {ICLR} 2026, Rio de Janeiro, Brazil, April 23-27, 2026},
  publisher    = {proceedings.iclr.cc},
  year         = {2026},
  url          = {https://proceedings.iclr.cc/paper\_files/paper/2026/hash/444a3737adaee10d86ad2ef5f74468e6-Abstract-Conference.html},
  bibsource    = {dblp computer science bibliography, https://dblp.org}
}

@inproceedings{SREGym,
  author       = {Jackson Clark and
                  Yiming Su and
                  Saad Mohammad Rafid Pial and
                  Lily Gniedziejko and
                  Tianyin Xu},
  title        = {SREGym: {A} Live Training Ground for {AI} {SRE} Agents with High-Fidelity
                  Failure Drills},
  booktitle    = {Proceedings of the {ACM} Conference on {AI} and Agentic Systems, {CAIS}
                  2026, San Jose, CA, USA, May 26-29, 2026},
  pages        = {1253--1258},
  publisher    = {{ACM}},
  year         = {2026},
  url          = {https://doi.org/10.1145/3786335.3813208},
  doi          = {10.1145/3786335.3813208},
  bibsource    = {dblp computer science bibliography, https://dblp.org}
}


\end{document}